\documentclass[12pt,journal,onecolumn,draftclsnofoot]{IEEEtran}
\usepackage{amsmath,amssymb,amsthm}
\usepackage[hidelinks]{hyperref}
\newtheorem{theorem}{Theorem}
\newtheorem{lemma}[theorem]{Lemma}
\newtheorem{proposition}[theorem]{Proposition}
\newtheorem{corollary}[theorem]{Corollary}
\newcommand{\Div}{\mathcal D}
\DeclareMathOperator{\wt}{wt}
\hypersetup{pdftitle={Optimal Cyclotomic Spectra for Distance-Preserving Extensions of Cyclic Codes},pdfauthor={Li Zhu and Hongfeng Wu}}
\title{Optimal Cyclotomic Spectra for Distance-Preserving Extensions \mbox{of Cyclic Codes}}
\author{Li~Zhu$^{1}$ and Hongfeng~Wu$^{2,*}$%
\thanks{$^{1}$School of Mathematical Sciences, Guizhou Normal University, Guiyang, China (e-mail: \href{mailto:izhumath@pku.edu.cn}{izhumath@pku.edu.cn}).}%
\thanks{$^{2}$College of Science, North China University of Technology, Beijing, China (e-mail: \href{mailto:whfmath@gmail.com}{whfmath@gmail.com}).}%
\thanks{$^{*}$Corresponding author: Hongfeng~Wu.}}
\begin{document}
\maketitle
\begin{abstract}
Let $n>1$ be coprime to a prime power $q$, and let $k$ be the number of distinct prime factors of $n$.
The length-$n$ cyclic code over $\mathbb F_q$ whose Fourier support consists of the characters of exact
order $n$ has minimum Hamming distance $2^k$. We determine how far its dimension can be increased by
passing to a cyclic supercode invariant under all multipliers while preserving this distance. A weighted
optimization on a finite divisor grid gives the maximum dimension and recovers the generator polynomials
of all maximizing codes. We also give an arithmetic criterion for attaining the dimension bound imposed
by the coordinate lines of the grid. For each fixed $t\ge0$ and every sufficiently large $h$, we construct squarefree lengths
with $h+t$ prime factors. In admissible characteristics dividing $h-1$, the possible sets of additional omissions
correspond precisely to minimum-cost coverings of $t$-subsets by $(t+1)$-subsets. In the other
admissible characteristics the additional omissions in the unique optimal spectrum are indexed by all
$t$-subsets.
Any prescribed finite nonempty set of primes can
occur as the set of admissible characteristics in which the maximum dimension is smaller than in
characteristic zero. For fixed $k$, we determine the first three characteristic-independent terms of the minimum redundancy
as the smallest prime power exactly dividing $n$ tends to infinity. When $k$ grows, two different
characteristics can give distinct positive limiting rates at the same lengths and minimum distances.
\end{abstract}

\section{Introduction}\label{sec:intro}

Let $C_0=\langle g_0(X)\rangle$ be a nonzero cyclic code of length $n$ over $\mathbb F_q$, where $g_0$ is
the monic generator polynomial of $C_0$. Each monic divisor $g$ of $g_0$ generates a cyclic code of the
same length containing $C_0$, and every cyclic supercode of $C_0$ arises in this way. Replacing $g_0$ by
$g$ increases the dimension by $\deg g_0-\deg g$. The minimum Hamming distance, on the other hand, may
decrease. It is therefore natural to ask how much the dimension of $C_0$ can be increased while both its
length and its minimum distance are kept fixed. Which divisors of $g_0$ give all the supercodes attaining
this maximum dimension?

Several constructions address the question of increasing dimension at a prescribed distance. Zhang, Cao,
and Luo~\cite{zcl} obtain a family of cyclic codes at certain even lengths $n\mid(q+1)$ in which the
dimensions increase while the minimum distance stays equal to $n/2$. Their Table~1 gives four codes of length $24$ with dimensions $2,4,6,8$ and common minimum distance $12$. Bernal, Bueno-Carre\~no,
and Sim\'on~\cite{bbs1} study maximum dimension at fixed length and fixed apparent distance for abelian
codes. Their binary example on the group $C_5\times C_7$ gives an inclusion of a $[35,27,4]$ code in a
$[35,28,4]$ code. In the bivariate presentation, the larger code has apparent distance four. It contains a word of weight four from the
smaller code~\cite[Example~37 in the authors' version]{bbs1}. Here $C_m$ denotes the cyclic group of
order $m$, so the Chinese remainder theorem identifies these codes with cyclic codes of length $35$. An
earlier construction of a binary cyclic $[35,28,4]$ code appears in van Lint and
Wilson~\cite[Example~4]{vlw}. For general linear codes, the supercode distances introduced
in~\cite[Definition~3.2]{cgl} record the largest minimum distance attainable at each prescribed
dimension. In the question above we prescribe the actual minimum distance of a given code and require its
supercodes to retain the cyclic structure.

In this paper we consider this question when $\gcd(n,q)=1$ and the generator polynomials are products of
complete cyclotomic factors. Write $\Phi_d$ for the $d$th cyclotomic polynomial over $\mathbb F_q$. We
start with the code
\[
 V_n^{(n)}=\left\langle (X^n-1)/\Phi_n(X)\right\rangle.
\]
For each proper divisor $d$ of $n$, we allow the entire factor $\Phi_d$ to be removed from the generator
polynomial. Such a factor may split further over $\mathbb F_q$. Keeping its irreducible factors together
amounts to requiring the supercode to be invariant under every unit multiplier on the coordinates, as we
recall in Section~\ref{sec:prelim}. The factor $\Phi_d$ accounts for $\varphi(d)$ dimensions, where
$\varphi$ is Euler's totient function. In particular, the starting dimension is $\varphi(n)$. Bhagat and
Sarma~\cite{bs1,bs2} studied the codes generated by $\Phi_n$ and $\Phi_n\Phi_1$ and their Euclidean
duals. In our notation, their distance results give $d(V_n^{(n)})=2^{\omega(n)}$ and show that, for
composite $n$, adjoining the order-one component preserves this distance. Here $\omega(n)$ is the number
of distinct prime divisors of $n$. When $n=pr$ is a product of distinct primes, their constant-component
result and the factorizations of $X^p-1$ and $X^r-1$ already give the unique optimal generator
$\Phi_p\Phi_r$. We give the short deduction in Section~\ref{sec:prelim}.

The effect on the distance depends on which factors are removed together. For example, take $n=30$ and
$\gcd(q,30)=1$. The starting code has dimension and minimum distance both equal to eight. Removing
$\Phi_1$ preserves its distance. Removing $\Phi_5$ alone also preserves it. Under the decomposition
$30=6\cdot5$, adjoining order five gives the tensor product of the length-six code on orders $1,6$ with
the length-five component of order five. Their distances are four and two, respectively. If both factors
are removed, however, the resulting code contains
\[
                    1+X^5+X^{10}+X^{15}+X^{20}+X^{25}.
\]
This word has weight six and its only nonzero Fourier coefficients occur at roots of orders one and five.
Thus the dimension gains from individual factors add, whereas their compatibility at distance eight has
to be decided jointly.

Let $M_q(n)$ denote the greatest dimension attainable in the class just described. Our first result
reduces its determination to an optimization on a grid obtained by truncating the prime-power exponents
of $n$. A point of the smaller grid represents a set of original orders. To omit this point one must omit
at least one order in the corresponding set. The smallest totient in that set is therefore the cost of
the omission. Theorem~\ref{cmp:minimum} specifies exactly which omissions preserve the prescribed
distance and how to lift every minimum-cost choice to the full list of optimal generator polynomials. The
maximum increase asked for above is then $M_q(n)-\varphi(n)$. For $k=\omega(n)\ge2$ the smaller grid has
at most $k^k$ points. Testing the distance condition and minimizing the total cost are further steps in
the computation.

Some lengths admit an explicit answer. Writing $P_i=p_i^{a_i}\parallel n$ for the distinct prime powers
exactly dividing $n$, we have
\[
 M_q(n)\le U(n):=n-\sum_{i=1}^k\varphi(n/P_i).
\]
Each summand is the least dimension that must be lost on one coordinate line of the divisor grid.
Theorem~\ref{sat:all} gives necessary and sufficient arithmetic conditions for equality and determines
the generators in all equality cases.

For each fixed integer $t\ge0$ and every sufficiently large integer $h$, Theorem~\ref{design:realization}
constructs a squarefree length $n=p_1\cdots p_k$ with $p_1<\cdots<p_k$ and $k=h+t$.
Write $p_B=\prod_{i\in B}p_i$ for $B\subseteq\{1,\ldots,k\}$, with $p_\varnothing=1$.
Beyond the compulsory omissions $n/p_i$, the unique optimum in characteristic zero or in prime
characteristic $\lambda\nmid n(h-1)$ omits all $p_B$ with $|B|=t$.
If $\lambda\nmid n$ and $\lambda\mid h-1$, the additional omitted orders are the $p_B$ for the
blocks of a minimum-cost covering of the $t$-subsets by $(t+1)$-subsets, where a block $B$ costs
$\prod_{i\in B}(p_i-1)$. Every such covering gives an optimal code. In this branch, the ordinary
covering number $C(k,t+1,t)$, the least possible number of blocks, can be recovered from the maximum
dimension by the formula
\[
 C(k,t+1,t)=\left\lfloor\frac{U(n)-M_q(n)}{(p_1-1)^{t+1}}\right\rfloor .
\]
A numerical illustration of the dependence on the characteristic appears after Corollary~\ref{gap:region}.

After establishing the notation and the starting distance in Section~\ref{sec:prelim}, we prove the
reduction and the finite distance test in Section~\ref{cmp:sec}. Sections~\ref{sec:saturation}
and~\ref{sec:gap} treat the coordinate-line bound and the covering designs, respectively. In
Section~\ref{sec:cubic} we determine three characteristic-independent terms of $n-M_q(n)$ for fixed $k$
and $\min_iP_i\to\infty$. Section~\ref{sec:persistent} treats a family on ten fixed prime factors and
then allows $k$ to grow. For any two distinct prime characteristics we construct a common sequence of
squarefree lengths on which the optimal rates converge to different positive limits. The common minimum
distance tends to infinity along this sequence, and its ratio to the length tends to zero.

\section{Preliminaries}\label{sec:prelim}

\subsection{Cyclic supercodes and their Fourier components}
Throughout the paper $q$ is a prime power and $n>1$ is coprime to $q$, unless another length is
specified. We identify a vector $c=(c_t)_{t\in\mathbb Z/n\mathbb Z}$ with the residue class of
$c(X)=\sum_{t=0}^{n-1}c_tX^t$ in $R_n=\mathbb F_q[X]/(X^n-1)$. Its support is
$\operatorname{supp}(c)=\{t:c_t\ne0\}$, and its Hamming weight is $\wt(c)=|\operatorname{supp}(c)|$. For
a nonzero linear code $C$ let $d(C)$ be the least weight of a nonzero word in $C$, and put $d(0)=\infty$.
A cyclic code is an ideal of $R_n$. For monic divisors $g,g_0$ of $X^n-1$ one has
\[
 \langle g_0\rangle\subseteq\langle g\rangle
 \quad\Longleftrightarrow\quad g\mid g_0,\qquad
 \dim\langle g\rangle=n-\deg g.
\]
Indeed, membership of a residue class in $\langle g\rangle$ is equivalent to divisibility of its
polynomial representative by $g$. Thus, if $g\mid g_0$, the distance is preserved precisely when $g$
divides no nonzero polynomial of degree less than $n$ having fewer than $d(\langle g_0\rangle)$ nonzero
coefficients. We use \emph{supercode} for a code in this same coordinate space, allowing equality.

Fix a primitive $n$th root of unity $\zeta$ in a splitting field. The characters of the additive group
$\mathbb Z/n\mathbb Z$ are $\chi_j(t)=\zeta^{jt}$. We use the Fourier coefficient $\widehat
c(j)=c(\zeta^j)=\sum_t c_t\chi_j(t)$. Its inverse transform has kernel $n^{-1}\zeta^{-jt}$. The Fourier
support $F(c)$ is the set of indices $j$ for which $\widehat c(j)\ne0$. The Fourier support of a code is
$F(C)=\bigcup_{c\in C}F(c)$, and its defining set $T(C)=(\mathbb Z/n\mathbb Z)\setminus F(C)$ indexes the
common zeros of its words. For $C=\langle g\rangle$, these are exactly the exponents of the roots of $g$.

Let $\Div(n)$ be the set of positive divisors of $n$. For $d\mid n$, put
\[
 E_d=\{j:\operatorname{ord}(\zeta^j)=d\}
     =\{j:n/\gcd(n,j)=d\}.
\]
Write $V_d^{(n)}$ for the unique cyclic code of length $n$ whose Fourier support is $E_d$. Since $X^n-1$
is squarefree, its cyclotomic factors are pairwise coprime over the base field. The Chinese remainder
theorem gives, for $S\subseteq\Div(n)$,
\begin{equation}\label{eq:code}
 \begin{aligned}
 V_d^{(n)}&=\left\langle (X^n-1)/\Phi_d(X)\right\rangle,\\
 D_S^{(n)}= D_S&=\bigoplus_{d\in S}V_d^{(n)}
       =\langle g_S(X)\rangle,\qquad
 g_S(X)=\prod_{d\in\Div(n)\setminus S}\Phi_d(X).
 \end{aligned}
\end{equation}
In particular $F(D_S)=\bigcup_{d\in S}E_d$, and $\dim D_S=\sum_{d\in S}\varphi(d)$. Reduction modulo
$\Phi_d$ identifies $V_d^{(n)}$ with $\mathbb F_q[X]/(\Phi_d)$ as an $\mathbb F_q$-vector space. The
polynomial $\Phi_d$ is the check polynomial of $V_d^{(n)}$. The projection of a word $c$ onto $V_d^{(n)}$
has the same Fourier coefficients as $c$ on $E_d$ and zero Fourier coefficients elsewhere. We call this
vector the \emph{order-$d$ component} of $c$. We call $S(c)=\{\operatorname{ord}(\zeta^j):j\in F(c)\}$
the \emph{order spectrum} of $c$. Its elements are exactly the orders with nonzero component. Thus $S$ is
the order spectrum of the code $D_S$. In particular, $\widehat c(0)=\sum_t c_t$ is a scalar, while the
order-one component is the constant vector $n^{-1}\widehat c(0)(1,\ldots,1)$. We refer to $V_n^{(n)}$ as
the \emph{primitive component}, and write $\delta_n(S)=d(D_S^{(n)})$, or $d(S)$ if the length is clear.

The unit multiplier $\mu_u$, $u\in(\mathbb Z/n\mathbb Z)^\times$, sends position $t$ to $ut$ and acts on
a word by $c(X)\mapsto c(X^u)\bmod(X^n-1)$. Hence $F(\mu_uc)=u^{-1}F(c)$. The $E_d$ are the orbits of the
full multiplier group since reduction of units modulo $n$ onto the units modulo $d$ is surjective. Every
cyclic code is determined by its defining set, so the codes in~\eqref{eq:code} are exactly the cyclic
codes invariant under every multiplier. By comparison, one irreducible factor of $\Phi_d$ corresponds to
one orbit of $j\mapsto qj$, a $q$-cyclotomic coset modulo $n$. For $d>1$, the set $E_d$ consists of
$\varphi(d)/\operatorname{ord}_d(q)$ such orbits, where $\operatorname{ord}_d(q)$ is the multiplicative
order of $q$ modulo $d$. For example, over $\mathbb F_2$ at length seven, $E_7=\{1,2,4\}\cup\{3,6,5\}$.
The two orbits correspond to the factors $X^3+X+1$ and $X^3+X^2+1$ of $\Phi_7$. Thus $V_7^{(7)}$ is the
direct sum of two minimal cyclic ideals of dimension three. The complete primitive component has
dimension six.

\subsection{Product coordinates and the starting distance}
Write $n=\prod_{i=1}^k P_i$, where $P_i=p_i^{a_i}$, the $p_i$ are distinct primes, and
$[k]=\{1,\ldots,k\}$. We distinguish two grids. The \emph{position grid} $\prod_i\mathbb Z/P_i\mathbb Z$
indexes the entries of a word, by the Chinese remainder theorem. The \emph{exponent grid}
$\prod_i\{0,\ldots,a_i\}$ indexes the divisors $d_e=\prod_i p_i^{e_i}$ of $n$. A \emph{slice} of a word
fixes a coordinate position. A \emph{section} of a spectrum fixes the exponent on an axis and retains the
allowed indices on the remaining axes. A coordinate face fixes some coordinates and lets the others vary
freely. A coordinate line has one free coordinate. The top index on axis $i$ is $a_i$. The vector with
top index on every axis corresponds to the order $n$. A position indicator takes value one at the
prescribed position and zero elsewhere.

For coprime lengths $m,h$, the product of characters identifies the order-$uv$ component at length $mh$
with $V_u^{(m)}\otimes V_v^{(h)}$ for $u\mid m$ and $v\mid h$. The identification descends to the base
field since the complete order classes are stable under Frobenius. For any two nonzero codes one has
$d(A\otimes B)=d(A)d(B)$. To see the lower bound, arrange a nonzero word as an array with columns in $A$
and rows in $B$. One nonzero column meets at least $d(A)$ nonzero rows, each of weight at least $d(B)$.
The tensor of two minimum-weight words attains the bound.

On a prime-power axis of length $p^a$, the sum of the layers of orders $1,p,\ldots,p^v$ consists of the
words periodic modulo $p^v$. This follows either from the Fourier transform or by repeating the
length-$p^v$ word in all residue periods. The top layer is the kernel of reduction modulo
$X^{p^{a-1}}-1$, which sums the coordinates within each residue class modulo $p^{a-1}$. It contains a
difference of two position indicators in the same class and has distance two. For $1\le e\le a$,
the order-$p^e$ layer is the $p^{a-e}$-fold repetition of the top layer at length $p^e$ and hence has
distance $2p^{a-e}$. In group-algebra terms, this is Proposition~2.1 of Dutra, Ferraz, and Polcino
Milies~\cite{dfm}, applied to the nested subgroups of orders $p^{a-e}$ and $p^{a-e+1}$ in the cyclic
group of order $p^a$. Taking tensor products across the prime-power axes gives $d(V_n^{(n)})=2^k$,
also a special case of their Proposition~2.3. We put
\begin{equation}\label{eq:objective}
 M_q(n)=\max_{\substack{n\in S\subseteq\Div(n)\\ \delta_n(S)=2^k}}
             \sum_{d\in S}\varphi(d).
\end{equation}
A subset in this maximum will be called a \emph{feasible spectrum}, and a maximizing subset an
\emph{optimal spectrum}. A Hamming isometry of codes preserves Hamming distance on words. The redundancy
of a length-$n$ code of dimension $K$ is $n-K$, so the minimum redundancy in~\eqref{eq:objective} is
$n-M_q(n)$.

For composite $n$, adjoining the constant component gives
$d(V_n^{(n)}\oplus V_1^{(n)})=2^k$~\cite[Th.~4.11]{bs2}. To see this, let $p$ be the least prime
divisor of $n$. Each primitive character sums to zero on each coset of the subgroup of order $p$.
Consequently, for $c\in V_n^{(n)}$ and $b\ne0$, the word $c+b\mathbf1$ has coset sums $pb\ne0$ and
weight at least $n/p$. This is at least $2^k$ except at $n=6$. For a composite prime power $n/p\ge2$,
for $k=2$ only $n=6$ gives $n/p<4$, and for $k\ge3$ one has $n/p\ge3^{k-1}\ge2^k$.
At length six the $2\times3$ array has column sums $2b$ and row sums $3b$. Weight at most three
would put one entry $2b$ in each column, leaving one row with sum $0$ or $2b$, a contradiction.
The primitive component supplies the matching upper bound. At prime length the sum of the two
components is the whole space.

\subsection{Comparison between coefficient fields}
An admissible prime characteristic at length $n$ is a prime $\lambda\nmid n$. The notation $M_0(n)$
denotes the same optimization over $\mathbb Q$. A set of positions contains the support of a nonzero word
if and only if the corresponding columns of a parity-check matrix are dependent. As matrix rank is
preserved by an extension of fields, the minimum distance is preserved by extending scalars. All
generator polynomials in~\eqref{eq:code} have coefficients in the prime field. Thus $M_q(n)$ depends on
$q$ only through $\lambda=\operatorname{char}\mathbb F_q$ and can also be denoted by $M_\lambda(n)$. The
proofs over characteristic zero use the same order components over $\mathbb Q$. We use $M_q(1)=1$,
$\omega(1)=0$ and the empty-product convention when no coordinate remains.

Two elementary examples will be useful for interpreting~\eqref{eq:objective}. Suppose $n=pr$ with $p$ and
$r$ distinct primes. The result just recalled allows the two orders $1,pr$ to be retained. On the other
hand,
\[
 V_{pr}^{(pr)}\oplus V_p^{(pr)}=\langle X^r-1\rangle,\qquad
 V_{pr}^{(pr)}\oplus V_r^{(pr)}=\langle X^p-1\rangle .
\]
Each of these supercodes contains a word of weight two. Hence the unique optimal spectrum is $\{1,pr\}$,
with generator $\Phi_p\Phi_r$ and dimension $M_q(pr)=(p-1)(r-1)+1$. In particular $M_2(35)=25$. The
defining set of the binary $[35,28,4]$ cyclic code of~\cite[Example~4]{vlw} is $\{5,7,10,14,20,21,28\}$.
It contains only indices of orders five and seven, and takes just one of the two binary cyclotomic cosets
of order seven. Consequently this code contains $V_{35}^{(35)}$, which supplies a word of weight four.
The defining set also gives the lower bound directly. In the $5\times7$ position grid, the parities of
the five rows agree, since the Fourier coefficients at all four roots of order five vanish. A word of
weight at most three has an empty row, so every row has even weight. A nonzero such word would have
two entries in the same row. Vanishing at the root $\zeta^5$ of order seven then forces their
positions to agree modulo seven as well. They would coincide in both coordinates. Thus the distance is
four, and the seven defining exponents give dimension $35-7=28$. Moreover, puncturing any
binary $[35,K,4]$ code gives a length-$34$ code of dimension $K$ and distance at least three, so the
disjoint Hamming balls of radius one give $35\cdot2^K\le2^{34}$ and hence $K\le28$. Requiring invariance
under all multipliers at this length and distance therefore costs exactly three dimensions.

\section{Truncation of the divisor grid and determination of the optimal generators}
\label{cmp:sec}

\subsection{Truncation in one prime-power coordinate}

To decide whether $D_S$ contains a word of weight less than a given
integer, how many of the exponents in a prime-power divisor of $n$ must
we distinguish? We first consider a single prime $p\mid n$. The relevant
operation on the spectrum is to intersect its sections at the smaller
powers of $p$. The following result gives the weight up to which this
operation preserves the distance test.

\begin{theorem}\label{cmp:block}
Let $n=mp^{b+s}$, where $p$ is a prime, $p\nmid m$, $b\ge0$, and $s\ge1$.
For $S\subseteq\Div(n)$, let
$S_u=\{d\in\Div(m)\mid dp^u\in S\}$ for $0\le u\le b+s$.
Define $A\subseteq\Div(mp^s)$ by prescribing its sections in the
$p$-coordinate as follows:
\[
 A_0=\bigcap_{u=0}^{b}S_u,\qquad
 A_j=S_{b+j}\quad(1\le j\le s).
\]
Then
\[
 \min\{2p^s,\delta_n(S)\}
   =\min\{2p^s,\delta_{mp^s}(A)\}.
\]
\end{theorem}

\begin{proof}
Put $L=p^s$. In the product of the coordinates of lengths $m$ and
$p^{b+s}$, write each position in the second coordinate uniquely as
$z+p^by$, with $0\le z<p^b$ and $0\le y<L$. Let $w_z$ be the block
of a word $w$ obtained by fixing $z$. We first determine when a word
supported in a single block belongs to $D_S$.

Fix $d\mid m$ and reduce the polynomial of $w$ modulo $\Phi_d$ in the
first coordinate. With coefficients in
the algebra $\mathcal A_d=\mathbb F_q[U]/(\Phi_d(U))$, the resulting polynomial has the form
\[
              \sum_{z=0}^{p^b-1} X^z W_z(X^{p^b}),\qquad
              W_z(Y)\in\mathcal A_d[Y],\quad \deg W_z<L
\]
for the nonzero $W_z$. If $j\ge1$, then
$\Phi_{p^{b+j}}(X)=\Phi_{p^j}(X^{p^b})$. Reduction modulo the
polynomial on the left therefore amounts to reducing every $W_z$
modulo $\Phi_{p^j}$. Indeed, after the latter reductions the terms
belonging to distinct $z$ have disjoint sets of exponents modulo $p^b$,
and the resulting polynomial has degree less than
$\deg\Phi_{p^{b+j}}$. Hence, for $w\in D_S$, every block satisfies
all the vanishing conditions specifying the positive sections of $A$.

If $0\le u\le b$, then $X^{p^b}=1$ modulo $\Phi_{p^u}(X)$.
Thus for a word supported in block $z$ the residue at level $u$ is
$X^zW_z(1)$. It vanishes precisely when $W_z(1)=0$, since $X$ is a
unit in the quotient and reduction by the monic $\Phi_{p^u}$ is
injective on the constant polynomials over $\mathcal A_d$. For this $d$, the
condition $W_z(1)=0$ is imposed exactly when at least one of the
sections $S_0,\ldots,S_b$ omits $d$. Taking all $d\mid m$, we see that
the word supported in block $z$ lies in $D_S$ if and only if the block
lies in $D_A$. In particular, $\delta_n(S)\le\delta_{mp^s}(A)$.

Suppose now that $D_S$ contains a nonzero $w$ of weight less than
both $2L$ and $\delta_{mp^s}(A)$. The preceding argument shows that
at least two blocks are nonzero. Choose a nonzero block $w_z$ of least
weight, and let $\sigma$ be the cyclic shift of its $y$-coordinate.
The difference $w_z-\sigma w_z$ still satisfies the conditions at
positive levels. Its sum over $y$ at each fixed first coordinate is
zero, so it satisfies all the conditions at level zero as well. It
belongs to $D_A$, and
\[
 \wt(w_z-\sigma w_z)\le 2\wt(w_z)\le\wt(w)<\delta_{mp^s}(A).
\]
Consequently $w_z=\sigma w_z$. Since this block is nonzero and
constant in $y$, its weight is at least $L$. Each other nonzero block
has at least the same weight, contradicting $\wt(w)<2L$.
\end{proof}

The cutoff $2p^s$ can be attained. To see this, take $b>0$ and fix
any spectrum $A$ at length $mp^s$. Set $S_0=A_0$,
$S_u=\Div(m)$ for $1\le u\le b$, and $S_{b+j}=A_j$ for $1\le j\le s$.
At a fixed position in the first coordinate, assign the value $1$
throughout one $y$-block and $-1$ throughout another, assigning zero
elsewhere. This word is periodic modulo $p^b$ and has zero sum in the
$p$-coordinate. Its only possible nonzero sections have indices
$1,\ldots,b$, all of which are allowed. Together with the embedding of
a single block, this gives
$\delta_n(S)\le\min\{2p^s,\delta_{mp^s}(A)\}$. By
Theorem~\ref{cmp:block} equality follows. In particular, if
$\delta_{mp^s}(A)>2p^s$, then the original distance is exactly $2p^s$.

\subsection{The minimum redundancy and all optimal generators}

Return to $n=\prod_{i=1}^k p_i^{a_i}$ and $D=2^k$. To test
whether a spectrum preserves $D$, choose the exponents
\[
 t_i=\min\{a_i,\max\{1,\lceil\log_{p_i}(D/2)\rceil\}\},\qquad
 b_i=a_i-t_i,\qquad n_*=\prod_{i=1}^k p_i^{t_i}.
\]
Thus, whenever the $i$th exponent is reduced, its cutoff
$2p_i^{t_i}$ is at least $D$. Write $d_u=\prod_i p_i^{u_i}$, and
identify each divisor with its exponent vector. The map to the
compressed exponent grid $G_*=\prod_i\{0,\ldots,t_i\}$ is
\[
           \rho(u)_i=\max\{0,u_i-b_i\}.
\]
For a retained spectrum $S$, put
\[
 H(S)=\{e\in G_*\mid \rho^{-1}(e)\not\subseteq S\},\qquad
 w_e=\prod_{i:e_i>0}\varphi(p_i^{b_i+e_i}),\qquad
 \boldsymbol t=(t_1,\ldots,t_k).
\]
A point $e$ belongs to $H(S)$ as soon as one divisor in its fiber has
been omitted. We next show why the smallest dimension loss for omitting
such a point is $w_e$.

\begin{theorem}\label{cmp:minimum}
For every $S\subseteq\Div(n)$, one has
\[
 \min\{D,\delta_n(S)\}
 =\min\{D,\delta_{n_*}(G_*\setminus H(S))\}.
\]
In particular, the minimum redundancy among the supercodes in
\eqref{eq:objective} is
\begin{equation}\label{cmp:opt}
 n-M_q(n)=
 \min_{\substack{H\subseteq G_*\setminus\{\boldsymbol t\}\\
                  \delta_{n_*}(G_*\setminus H)\ge D}}
                 \sum_{e\in H} w_e .
\end{equation}
For each minimizing $H$ and each $e\in H$, select a divisor
$d(e)\in\rho^{-1}(e)$ with $\varphi(d(e))=w_e$. Then the polynomials
\[
                      \prod_{e\in H}\Phi_{d(e)}(X)
\]
are precisely the monic generators of the supercodes attaining
$M_q(n)$, as $H$ and the choices $d(e)$ vary.
\end{theorem}

\begin{proof}
Apply Theorem~\ref{cmp:block} successively to the coordinates with
$b_i>0$, truncating both sides of each equality at $D$. At a zero
index in the new grid, the theorem takes an intersection of sections.
At a positive index there is only one original section. After the last
coordinate, a point $e$ is therefore retained if and only if all the
divisors of its fiber are retained. This proves the distance identity.

The fiber of $\boldsymbol t$ consists only of $n$, so its omission is
forbidden. For every other $e$, the indices with $e_i>0$ force
$u_i=b_i+e_i$ for each $u\in\rho^{-1}(e)$. In the remaining
coordinates one may take $u_i=0$, with totient contribution $1$. Thus
the least value of $\varphi(d_u)$ on this fiber is $w_e$.
For a prescribed $H$, one has to omit at least one divisor in every
fiber over $H$ and none elsewhere. Since the fibers are disjoint and
each omitted divisor has positive cost, the minimum loss is
$\sum_{e\in H}w_e$. It is attained exactly by omitting a
single divisor of least totient in each such fiber. The generators
now follow from~\eqref{eq:code}. The primitive component itself has
distance $D$, so at least one feasible choice exists.
\end{proof}

\subsection{An integer test for a prescribed set of omissions}

For $k\ge2$ one has $t_i\le k-1$, and hence $|G_*|\le k^k$.
For $k=1$ the compressed grid has two points. This count bounds the
number of points from which $H$ is selected. To evaluate
\eqref{cmp:opt}, it remains to test the distance condition for a
candidate $H$ and to minimize its total cost. We give a finite test
which also describes the dependence of the distance condition on the
characteristic.

\begin{proposition}\label{cmp:minor}
For each $H\subseteq G_*$ there is an explicitly constructible integer
$b_H\ge0$ such that, in every admissible positive characteristic
$\lambda$,
\[
          \delta_{n_*}(G_*\setminus H)\ge D
          \quad\Longleftrightarrow\quad \lambda\nmid b_H.
\]
In characteristic zero the condition is $b_H\ne0$.
Up to a permutation of the coordinates, $b_H$ depends only on $H$,
the pairs $(p_i,t_i)$ with $p_i<D$, and the number of remaining
coordinates. Every admissible characteristic greater than
$(D-1)^{(D-1)/2}$ has the same feasible spectra as characteristic zero.
\end{proposition}

\begin{proof}
Write $h=D-1$. In the $i$th coordinate with $p_i<D$, put
$T_i=\{0,\ldots,p_i^{t_i}-1\}$. For $0\le j\le t_i$, take the
columns of an integer matrix $L_{i,j}$ to be the coefficient vectors
of $X^x\bmod\Phi_{p_i^j}(X)$ in the monomial basis, for $x\in T_i$.
In particular, $L_{i,0}$ is the row of ones. At a positive level,
first reduce $x$ modulo $p_i^j$ and use
\[
            \Phi_{p_i^j}(X)=\sum_{v=0}^{p_i-1}X^{v p_i^{j-1}}.
\]
This expresses each residue vector with entries in $\{0,1,-1\}$.

In the $i$th coordinate with $p_i\ge D$ one has $t_i=1$.
A support of size at most $h$ uses at most $h<p_i$ distinct
positions in this coordinate. The corresponding monomials are
linearly independent modulo $\Phi_{p_i}$, because the only relation
among all $p_i$ monomials, up to a scalar, is their sum. Thus these
positions can be relabelled within $T_i=\{1,\ldots,h\}$. On this
label set it suffices to take $L_{i,0}$ to be a row of ones and
$L_{i,1}$ to be the identity matrix of size $h$.

Stack the matrices $\bigotimes_i L_{i,e_i}$ for $e\in H$ to form
$Q_H$ with column set $T=\prod_i T_i$. For any support of size at
most $h$ in the original length $n_*$, each relabelling above gives
a linear isomorphism on the spans of the level-one column vectors and
preserves the level-zero row. Tensoring these identifications preserves
the kernel on the chosen support at each omitted order, and hence their
common kernel. Conversely, each $T_i$ can be embedded in the actual
coordinate set, so a dependence among at most $h$ columns of $Q_H$
also gives a word of weight at most $h$ at length $n_*$. It follows
that the desired distance condition is precisely the independence of
every set of at most $h$ columns of $Q_H$.

It suffices to use sets of exactly $h$ columns. Indeed, if some
$p_i\ge D$, then $|T|\ge h$. If all $p_i<D$, then $|T|=n_*\ge2^k$.
For $U\subseteq T$ of size $h$, let $g_U$ be the greatest common
divisor of the absolute values of all $h\times h$ minors on the
columns in $U$. Set $g_U=0$ if there are no such minors or all are
zero. In positive characteristic $\lambda$ the columns in $U$ are
independent precisely when $\lambda\nmid g_U$, and in characteristic
zero precisely when $g_U\ne0$. Hence one may take
\[
                         b_H=\prod_{|U|=h}g_U.
\]
For the coordinates satisfying $p_i\ge D$, the same two matrices are
used independently of the prime. This proves the asserted dependence
of $b_H$ on the data. Finally, Hadamard's inequality gives
$|\det B|\le h^{h/2}$ for each $h\times h$ submatrix $B$ of $Q_H$,
since its entries are $0$, $1$, or $-1$. Above this bound every
nonzero minor stays nonzero after reduction, which proves the last
assertion.
\end{proof}

In particular, suppose $n\in S$ and $d(D_S)=D$. Given
$E\subseteq\Div(n)\setminus S$, the factors $\Phi_d$, $d\in E$, can
all be removed from the generator of $D_S$ while preserving its
distance if and only if $\lambda\nmid b_{H(S\cup E)}$. When this
condition holds, the resulting dimension gain is
$\sum_{d\in E}\varphi(d)$.

Every spectrum feasible in positive characteristic is feasible over
$\mathbb Q$, since a minor nonzero modulo $\lambda$ is already nonzero
as an integer. Consequently $M_q(n)\le M_0(n)$, with equality in all
sufficiently large admissible characteristics. An admissible prime
$\lambda$ for which $M_\lambda(n)<M_0(n)$ will be called
\emph{exceptional at length $n$}.

\section{An arithmetic criterion for attaining the coordinate-line bound}\label{sec:saturation}

Let $n=\prod_{i=1}^kP_i$, where $P_i=p_i^{a_i}$ and $p_1<\cdots<p_k$.
We first consider a necessary condition on the factors of the generator polynomial of a distance-preserving supercode.
Fix $i$. In the exponent grid, the line through $(a_1,\ldots,a_k)$ in direction $i$ consists of the
orders $(n/P_i)p_i^e$, $0\le e\le a_i$.
If the whole line were retained, the code would contain all tensors obtained by taking an arbitrary word
on the $i$th axis and a word in the highest-order component on each other axis.
Taking a unit vector on the free axis and a difference of two unit vectors on each of the other axes
gives a word of weight $2^{k-1}$.
Thus at least one order on the line must be omitted. The $k$ lines meet only at $n$, which is retained,
and the least dimension lost on line $i$ is $\varphi(n/P_i)$. Put
\[
 U(n)=n-\sum_{i=1}^k\varphi(n/P_i),\qquad
 B_n=\Div(n)\setminus\{n/P_i:1\le i\le k\}.
\]
The preceding argument gives $M_q(n)\le U(n)$. The following theorem determines exactly when the omissions defining $B_n$ suffice.

\begin{theorem}\label{sat:all}
For $k\ge2$, the equality $M_q(n)=U(n)$ holds if and only if the following three conditions hold simultaneously.
\begin{align}
 2p_i&\ge 2^k &&\text{for each $i$ with }a_i\ge2, \label{sat:repeat}\\
 p_1p_2&\ge 2^k, \label{sat:plane}\\
 \sum_{i=1}^h(p_i-1)&\ge 2^h &&(3\le h\le k). \label{sat:prefix}
\end{align}
For $k=1$ we have $M_q(n)=U(n)=n-1$. In every case of equality, one maximizing spectrum is $B_n$, with generator polynomial
\[
 g_B(X)=\prod_{i=1}^k\Phi_{n/P_i}(X).
\]
It is the unique maximizing spectrum if $4\nmid n$. If $4\mid n$, there is exactly one further maximizing
spectrum. Its generator polynomial is obtained from $g_B(X)$ by replacing $\Phi_{n/P_1}(X)$ by
$\Phi_{2n/P_1}(X)$.
\end{theorem}

We shall use two facts about words of small weight on a grid of coordinate positions. The first applies
to arbitrary finite sets of positions, without requiring a cyclic structure on each axis.
Let $j\ge1$ and let $X_1,\ldots,X_j$ be nonempty finite sets, with $b_i=|X_i|$ and
$1\le b_1\le\cdots\le b_j$.
On $\prod_{i=1}^j X_i$, write $\mathcal L_u$ for the linear span over the coefficient field of functions
depending on at most $u$ coordinates. A coordinate flat is the subset obtained by fixing the positions on
a specified set of axes and allowing the other positions to vary.

\begin{lemma}\label{gap:flat}
For $0\le u\le j$, the minimum distance of $\mathcal L_u$ is $P=\prod_{i=1}^{j-u}b_i$.
Suppose in addition that $b_1\ge2$ and $b_{i+1}\ge b_i+2$ for all $i<j$.
Every nonzero word of weight less than $2P$ in $\mathcal L_u$ is then a scalar multiple of the indicator
of a coordinate flat with $j-u$ free axes.
\end{lemma}

\begin{proof}
If $b_j=1$, the grid consists of a single position and $d(\mathcal L_u)=1$ for every $u$. Assume
$b_j\ge2$. The cases $u=0$ and $u=j$ give the repetition code and the full space, respectively.
For $0<u<j$, fix the last coordinate at two different positions. Each of the
resulting slices lies in $\mathcal L_u$ on the first $j-1$ axes. Their difference lies in $\mathcal
L_{u-1}$ on these axes. Indeed, a summand independent of the last coordinate cancels, and each remaining
summand involves at most $u-1$ of the first $j-1$ coordinates.
By induction the respective minimum distances on the smaller grid are $P'=P/b_{j-u}$ and $P$. If all
slices are nonzero, their total weight is at least $b_jP'\ge P$. If some slice is zero, every nonzero
slice is a slice difference and has weight at least $P$. The indicator of a flat whose free axes have
sizes $b_1,\ldots,b_{j-u}$ attains $P$.

Now assume the separation of the side lengths and consider a nonzero word of weight less than $2P$.
If all slices coincide, their common weight is less than $2P/b_j\le 2P'$. The induction hypothesis gives
the required description, with the last axis free.
If one slice is zero, exactly one can be nonzero. Applying the induction hypothesis to this slice in
$\mathcal L_{u-1}$ gives a flat with the last coordinate fixed.
In the remaining case choose two unequal slices. Their weights add to at least $P$, and each of the
remaining $b_j-2$ slices has weight at least $P'$. Since $j-u<j$, the separation assumption gives
$b_j-2\ge b_{j-u}$. The total weight would therefore be at least $P+(b_j-2)P'\ge2P$.
\end{proof}

The distance formula in Lemma~\ref{gap:flat} may also be obtained by interpolation from the distance of
decreasing monomial-Cartesian codes~\cite[Th.~3.9]{clms}.

We next take the side lengths to be distinct primes. For a set $J$ of indices, write $n_J=\prod_{i\in
J}p_i$, $s_i=p_i-1$, and $L(I)=\sum_{i\in I}s_i$ for $I\subseteq J$.
Each divisor of $n_J$ specifies the axes on which a character is nontrivial. In particular, the
characters of order $n_J/p_i$ are constant on axis $i$ and nontrivial on each other axis.
For a position $x$ in the grid, let $\ell_i(x)$ be the full line in direction $i$ through $x$.
Subtracting from the unit vector at $x$ its coordinate average along each such line gives the word
\[
 Z_{J,x}=\mathbf1_{\{x\}}-\sum_{i\in J}p_i^{-1}\mathbf1_{\ell_i(x)} .
\]
Here $p_i^{-1}$ is taken in the coefficient field. At $x=0$, the corresponding polynomial satisfies, for $d\mid n_J$,
\begin{equation}\label{gap:congruence}
 \begin{aligned}
 Z_{J,0}(X)
 &=1-\sum_{i\in J}p_i^{-1}\sum_{r=0}^{p_i-1}X^{r n_J/p_i}\\
 &\equiv 1-|J|+\omega(d)\pmod{\Phi_d(X)} .
 \end{aligned}
\end{equation}
Indeed, at a primitive $d$th root the inner sum is $p_i$ when $p_i\nmid d$, and is zero when $p_i\mid d$.
Translation preserves the order spectrum, so this evaluation also determines the orders occurring in
$Z_{J,x}$. The lines contribute exactly $L(J)$ nonzero entries away from $x$. At $x$ the entry is
$1-\sum_{i\in J}p_i^{-1}$, which can vanish in the coefficient field. Thus $\wt(Z_{J,x})$ is $L(J)$ or
$L(J)+1$.

\begin{lemma}\label{gap:star}
Let $j=|J|\ge2$ and $S\subseteq\Div(n_J)$. Assume that $n_J/p_i\notin S$ for all $i\in J$ and put $I=\{i\in J:p_i\ge5\}$.
Let $w\in D_S$. Suppose that in each direction $i\in I$ every nonzero difference of two slices of $w$ has
weight at least $2^{j-1}$. If $\wt(w)<2^j$ and the order-$n_J$ component of $w$ is nonzero, then
$\wt(w)\ge L(I)$.
If in addition $I=J$ and
\[
 \wt(w)<L(J)+\min_{i\ne r\in J}(s_i s_r-s_i-s_r),
\]
then $w$ is a nonzero scalar multiple of $Z_{J,x}$ for some $x$.
\end{lemma}

\begin{proof}
For each $i\in I$ choose a slice of least weight, say $b$, in direction $i$.
Any slice unequal to the chosen one has weight at least $2^{j-1}-b$.
If there were at least two such slices, summing their weights and the weights of the remaining slices would give
\[
 \wt(w)\ge2(2^{j-1}-b)+(p_i-2)b\ge2^j .
\]
On the other hand, equality of all slices would make $w$ constant in direction $i$ and force its
order-$n_J$ component to vanish. Hence the chosen slice is repeated at each position except one, which we
denote by $x_i$.

Let $E_i=\mathbf1_{\{x_i\}}$ on axis $i$. In every direction in $I$ the word is a linear combination of
$1$ and $E_i$, so there is a unique expansion
\[
 w=\sum_{A\subseteq I} c_A\prod_{i\in I\setminus A}E_i .
\]
Here each $c_A$ is a function on the axes in $J\setminus I$. On the set of positions where $t_i\ne x_i$
precisely for $i\in T\subseteq I$, the value as a function of the remaining coordinates is
$v_T=\sum_{A\supseteq T}c_A$.

Write $Q$ for projection onto the highest-order component on the axes in $J\setminus I$, taking $Q$ to be
the identity on scalars when $J=I$.
In each direction in $I$, $E_i$ has constant component $p_i^{-1}\mathbf1$ and nonzero complementary
component. Projection onto order $n_J$ therefore gives $Qc_\varnothing\ne0$. Projection onto order
$n_J/p_i$, which is omitted, gives
\[
 Q(c_{\{i\}}+p_i^{-1}c_\varnothing)=0 \qquad (i\in I).
\]
In particular $c_{\{i\}}\ne0$ for each $i\in I$.
By M\"obius inversion, $c_{\{i\}}=\sum_{T\supseteq\{i\}}(-1)^{|T|-1}v_T$.
It follows that the sets $T$ for which $v_T\ne0$ cover $I$.

Each such $T$ contributes at least $\prod_{i\in T}s_i$ to the weight. Counting the $s_i$ over this cover gives
\[
 \wt(w)\ge L(I)+
 \sum_{\substack{T\subseteq I,\ |T|\ge2\\v_T\ne0}}
 \left(\prod_{i\in T}s_i-\sum_{i\in T}s_i\right).
\]
Since $s_i\ge4$ on $I$, every summand is nonnegative. For $i\ne r$ in $T$ it is at least
$s_is_r-s_i-s_r$, since adjoining a further factor of size at least four can only increase the
difference. This proves the first assertion.
Under the further assumptions, $v_T$ must vanish whenever $|T|\ge2$. Inversion then gives $c_A=0$ for
$|A|\ge2$, and the projection equations give $c_{\{i\}}=-c_\varnothing/p_i$. Substitution in the
expansion of $w$ yields $w=c_\varnothing Z_{J,x}$.
\end{proof}

\begin{proof}[Proof of Theorem~\ref{sat:all}]
On the $i$th line of the exponent grid the cost of omitting $(n/P_i)p_i^e$ is
$\varphi(n/P_i)\varphi(p_i^e)$. Attaining $U(n)$ therefore requires one least-cost omission on each line
and no other omissions. The exponent $e=0$ always minimizes the cost. The only other possibility is
$p_1=2$, $e=1$ and $a_1\ge2$, since the top order must be retained. In that case multiplication in each
coordinate position by the character $t\mapsto(-1)^{t_1}$ interchanges orders one and two on the first
axis and preserves each higher-order layer there. It induces a Hamming isometry between the codes defined
by the two candidate spectra. It is consequently enough to determine when $D_{B_n}$ has distance $2^k$.
If $k=1$, $D_{B_n}$ is the zero-sum code and has distance two.

Suppose $k\ge2$. If $a_i\ge2$, then $B_n$ contains the entire section at exponent $a_i-1$ on axis $i$. At
length $p_i^{a_i-1}$ take a difference of two unit vectors in the same class modulo $p_i^{a_i-2}$, and
repeat it $p_i$ times to length $P_i$. It belongs to the layer of order $p_i^{a_i-1}$. Tensoring with a
unit vector on each remaining axis gives a word in $D_{B_n}$ of weight $2p_i$. Thus \eqref{sat:repeat} is
necessary. If \eqref{sat:repeat} holds, Theorem~\ref{cmp:block} can be applied on every repeated prime
axis with the new exponent equal to one. Its cutoff $2p_i$ is at least $2^k$. On each reduced grid the
omitted orders again have one zero index and maximal indices elsewhere. This reduces the remaining
distance test, truncated at $2^k$, to $B_{\prod_i p_i}$.

Suppose henceforth that $n$ is squarefree. The indicator of a coordinate plane with the first two axes
free has order spectrum contained in $B_n$ and weight $p_1p_2$. This gives \eqref{sat:plane}.
When $p_1\ge5$, the integers $s_i=p_i-1$ are even. If $L([h])<2^h$ for some $3\le h\le k$, then
$L([h])+1<2^h$. On the first $h$ axes take the word $Z_{[h],0}$ and tensor it with a nonzero two-point
difference on each remaining axis. The factors outside $[h]$ are in the highest layers
and~\eqref{gap:congruence} shows that the orders with just one constant direction in $[h]$ do not occur.
This gives a nonzero word of $D_{B_n}$ of weight less than $2^k$, proving \eqref{sat:prefix} in this
case.
If $p_1\le3$ and $k\ge3$, condition~\eqref{sat:plane} already implies
\begin{equation}\label{sat:smallside}
 \sum_{i=2}^h s_i\ge2^h\qquad(3\le h\le k).
\end{equation}
Indeed $p_2\ge5$. For $h=3,4$ the respective lower bounds $4+6$ and $4+6+10$ suffice. For $h\ge5$ use
$p_2\ge2^h/3$ and $(h-1)(2^h/3-1)\ge2^h$.

We prove sufficiency on the squarefree grid by induction on $k$. When $k=2$ we have $B_n=\{1,n\}$,
and the constant-component calculation in Section~\ref{sec:prelim} gives $d(D_{B_n})=4$.

For $k\ge3$, deleting any axis leaves a grid satisfying the same hypotheses at threshold $2^{k-1}$.
Indeed, the smallest prime products and the prefix sums on the smaller grid are bounded below by the
corresponding quantities for the original grid.
Fix $i$ and take two slices of a word $w\in D_{B_n}$ perpendicular to this axis. Their difference has order spectrum contained in
\[
 S_i^+=\{d\mid n/p_i:p_i d\in B_n\}=B_{n/p_i}.
\]
To see this, the terms constant on axis $i$ cancel, whereas evaluation and subtraction on this axis leave
each order in the other directions unchanged. In particular the orders $(n/p_i)/p_r$, $r\ne i$, are
missing, since they would come from the omitted orders $n/p_r$. These operations are defined over the
coefficient field. Hence every nonzero slice difference has weight at least $2^{k-1}$ by induction.

If the order-$n$ component of $w$ vanishes, all its remaining components are constant in at least two
directions. Thus $w\in\mathcal L_{k-2}$ and Lemma~\ref{gap:flat} gives $\wt(w)\ge p_1p_2$ for nonzero
$w$. Otherwise suppose that $\wt(w)<2^k$. Applying Lemma~\ref{gap:star} with $J=[k]$ gives
$\wt(w)\ge\sum_{i:p_i\ge5}s_i$. For $p_1\ge5$ this is at least $2^k$ by~\eqref{sat:prefix} with $h=k$.
For $p_1\le3$ it follows from~\eqref{sat:smallside}. Both cases contradict the assumed weight. Finally
$D_{B_n}$ contains the primitive component, whose minimum distance is $2^k$, so its distance equals $2^k$
as required.
\end{proof}

\subsection{Optimal generators at lengths 30 and 12}
We return to the example of length $30$ in the introduction. Here $U(30)=16$, and Theorem~\ref{sat:all} excludes equality because $2\cdot3<8$. The three coordinate lines force the orders $15,10,6$ to be omitted. Their totients sum to
$14$. To obtain a code of dimension $15$ we must omit precisely one further order of totient one, namely
$1$ or $2$. If $2$ is omitted, both $1$ and $5$ remain, and the six-term word in the introduction belongs
to the code. If $1$ is omitted, the remaining orders are $2,3,5,30$. Indeed, $30$ is squarefree, so $h(e)=1$ for every order. Under~\eqref{cubic:mask} with $z(e)\le2$,
the allowed numbers of zero indices are precisely zero and two. These give $30$ and the three prime
orders, respectively. The construction in the proof of Corollary~\ref{asy:expansion} therefore gives
minimum distance eight for this spectrum. Hence
$M_q(30)=15$, and the unique generator of a maximizing code is
\[
                         g(X)=\Phi_1(X)\Phi_6(X)\Phi_{10}(X)\Phi_{15}(X).
\]
In particular, a choice that starts by adjoining order one can be extended to a feasible spectrum maximal
under inclusion, but every such choice has dimension strictly less than $15$.

At length $12$ the same optimization illustrates the relation with the arithmetic Singleton bound. Write
$g_6=\Phi_2\Phi_3$, the generator of the $[6,3,4]_q$ code of Section~\ref{sec:prelim}. At length $12$ the
code generated by $g_6(X^2)=\Phi_3\Phi_4\Phi_6$ has parameters $[12,6,4]_q$. Its even and odd positions
give two independent copies of the length-six code, as is seen by writing each multiplier of $g_6(X^2)$
as $a(X^2)+Xb(X^2)$. Removing $\Phi_6$ gives the $[12,8,4]_q$ code generated by $g=\Phi_3\Phi_4$, which
attains $M_q(12)$ by Theorem~\ref{sat:all}. More generally, let $f$ be the monic generator of a nonzero cyclic code of length $n$,
with $0<\deg f<n$, and let $s$ be the greatest common divisor of the positive exponents with nonzero
coefficient in $f$. The arithmetic Singleton bound in this case is $d\le1+\deg(f)/s$~\cite{zwbound}.
Indeed, $f$ represents a nonzero codeword and every exponent with a nonzero coefficient belongs to
$\{0,s,\ldots,\deg f\}$, so $d\le\wt(f)\le1+\deg(f)/s$. For $g_6(X^2)$ we have
$s=2$ and the bound equals four. For $g$, the coefficient of $X$ is one, so $s=1$ and the bound equals
five. Thus increasing the dimension from six to its maximum of eight preserves the minimum distance,
while equality in the arithmetic Singleton bound is lost.

\section{Covering designs and exceptional characteristics}\label{sec:gap}

For a squarefree length the dimensions of the individual order components are
the same over all admissible fields. To see how the optimal dimension can
nevertheless depend on the field, we examine which components of a word of
small weight remain nonzero in a given characteristic.

Throughout this section let $n=\prod_{i=1}^k p_i$, where
$5\le p_1<\cdots<p_k$ are primes and $k\ge2$. For $A\subseteq[k]$ put
$p_A=\prod_{i\in A}p_i$, with $p_\varnothing=1$. An order with constant
directions $A$ is $n/p_A$. If $S$ is the set of retained orders, write
\[
 \mathcal H(S)=\{A\subseteq[k]:n/p_A\notin S\},\qquad
 w(A)=\varphi(n/p_A)=\prod_{i\notin A}s_i,
\]
where $s_i=p_i-1$. Thus $w(A)$ is the dimension lost by omitting that
order. Recall that $L(J)=\sum_{i\in J}s_i$, and put $L_h=L([h])$ and
$\Delta=s_1s_2-s_1-s_2$. Write $\lambda$ for the characteristic.
When $\lambda=0$ the coefficient field is $\mathbb Q$.

\begin{theorem}\label{design:covercriterion}
Suppose $n\in S$,
\[
 p_1p_2\ge2^{k-1},\qquad L_h+\Delta\ge 2^h\quad(2\le h\le k).
\]
Then $d(D_S)=2^k$ if and only if the following conditions hold.
\begin{enumerate}
\item $\{i\}\in\mathcal H(S)$ for each $i\in[k]$.
\item If $i<j$ and $p_ip_j<2^k$, at least one divisor of
$n/(p_ip_j)$ is omitted from $S$.
\item For every $J\subseteq[k]$ with $|J|\ge2$ and
$L(J)<2^{|J|}$, there is an $A\in\mathcal H(S)$ such that
$A\subseteq J$ and $1-|A|$ is nonzero in the coefficient field.
\end{enumerate}
If $s_1s_2\ge2^k$, the numerical hypotheses hold and condition~2 is automatic.
\end{theorem}

\begin{proof}
For necessity, the coordinate-line bound at squarefree length forces the
omission of each order $n/p_i$. For a coordinate plane with directions
$i,j$ free, its indicator has weight $p_ip_j$ and order spectrum
$\Div(n/(p_ip_j))$. Indeed, it is constant on each free axis and is a
point indicator on each remaining axis, with nonzero projection onto both
local components. This proves the first two conditions.

Now fix $J$ as in condition~3. Since every $s_i$ is even,
$L(J)<2^{|J|}$ implies $L(J)+1<2^{|J|}$. Tensor the word
$Z_{J,0}$ with a difference of two unit vectors on each axis outside
$J$. The resulting nonzero word has weight at most
$2^{k-|J|}(L(J)+1)<2^k$. The factors outside $J$ have only their
top components. By~\eqref{gap:congruence} the constant directions of
the nonzero components of the tensor are exactly the sets
$A\subseteq J$ for which $1-|A|\ne0$ in the field. To exclude this
word the code must omit at least one of these components.

Conversely, suppose conditions~1--3 hold. We prove a statement for all
nonempty subsets $J\subseteq[k]$ simultaneously. Put $n_J=p_J$ and
\[
 S_J=\{d\mid n_J : d\,n/n_J\in S\}.
\]
This is the section of $S$ obtained by fixing the indices outside $J$
at their top levels. We claim $d(D_{S_J})\ge2^{|J|}$, where $D_{S_J}$
has length $n_J$. For $|J|=1$, the top component is present and
condition~1 excludes the constant component. The claim follows from
the distance of the zero-sum code.

Assume the claim for smaller nonempty subsets and write $j=|J|\ge2$.
Let $w$ be a nonzero word of $D_{S_J}$. If its top component vanishes,
every remaining component of $w$ is constant in at least two directions,
by condition~1. Hence $w\in\mathcal L_{j-2}$. When $J\ne[k]$,
Lemma~\ref{gap:flat} gives
$\wt(w)\ge p_1p_2\ge2^{k-1}\ge2^j$. When $J=[k]$, we have
$2^k\le2p_1p_2$. The primes are odd and distinct, so the separation
hypothesis of the same lemma holds. A word of weight less than $2^k$
would be a scalar multiple of the indicator of a plane of weight
$p_i p_{i'}<2^k$, contrary to condition~2.

It remains to consider a word $w$ with nonzero top component. Differences
of two slices in direction $i\in J$ lie in $D_{S_{J\setminus\{i\}}}$:
the terms constant in direction $i$ cancel, and the other terms have
the top index on this axis. Thus each nonzero slice difference has
weight at least $2^{j-1}$. Also
\[
 L(J)+\min_{i\ne r\in J}(s_i s_r-s_i-s_r)\ge L_j+\Delta\ge2^j.
\]
If $\wt(w)<2^j$, Lemma~\ref{gap:star} gives
$w=cZ_{J,x}$ for some $c\ne0$ and some point $x$. Its weight is at
least $L(J)$, so condition~3 applies. It supplies an $A\subseteq J$
for which $n_J/p_A\notin S_J$ and the projection of $Z_{J,x}$ at
that order is nonzero, by~\eqref{gap:congruence}. This is impossible
for $w\in D_{S_J}$. The induction gives the lower bound at $J=[k]$,
and $V_n^{(n)}\subseteq D_S$ gives equality. Finally
$L_h+\Delta\ge s_1s_2$ for $h\ge2$, while every $p_i p_j$
with $i\ne j$ exceeds $s_1s_2$. This proves the last assertion.
\end{proof}

We next choose the primes so that the third condition of
Theorem~\ref{design:covercriterion} reduces to a classical covering
problem. A covering of the $t$-subsets of $[k]$ by $b$-subsets is a
collection of $b$-subsets, called blocks, such that every $t$-subset
lies in at least one block. The least possible number of blocks is the
covering number $C(k,b,t)$~\cite{cover}. For $t=0$ the empty set is the
one subset to be covered, so $C(k,1,0)=1$. In the following theorem
the cost of a block is $c(B)=\prod_{i\in B}s_i$. We shall need both
the number of blocks and the sum of their costs.

\begin{theorem}\label{design:realization}
Fix an integer $t\ge0$. For every sufficiently large integer $h$, let
$k=h+t$. There exist primes $p_1<\cdots<p_k$ with
$p_1\asymp 2^h/h$ such that for $n=p_1\cdots p_k$ the following hold over every admissible
coefficient field, including characteristic zero.
\begin{enumerate}
\item If $1-h$ is nonzero in the coefficient field, the unique
maximizing spectrum omits the $k$ orders $n/p_i$ and all orders $p_B$
with $|B|=t$. Its additional redundancy is
\[
 U(n)-M_q(n)=\sum_{|B|=t}c(B).
\]
\item If the characteristic divides $h-1$, the maximizing spectra omit
the $k$ orders $n/p_i$ and the orders $p_B$ for all blocks $B$ of a
covering of the $t$-subsets by $(t+1)$-subsets of minimum total cost.
Every such covering gives a maximizing spectrum, and every such covering
has $C(k,t+1,t)$ blocks. For these choices of $n$ and characteristic,
\begin{equation}\label{design:recover}
 C(k,t+1,t)=\left\lfloor\frac{U(n)-M_q(n)}{s_1^{t+1}}\right\rfloor .
\end{equation}
\end{enumerate}
In both cases all other orders are retained.
\end{theorem}

\begin{proof}
Put $x=2^h/h$. By the prime number theorem~\cite[Sec.~6.2]{mv},
the interval $[2x/3,3x/4]$ contains asymptotically $x/(12\log x)$
primes. Divide it into $\lceil h^{t+3}\rceil$ equal subintervals.
For sufficiently large $h$ one of these contains at least $k=h+t$
primes. Choose $p_1<\cdots<p_k$ there. Since $p_1\asymp x$ and
the relative length of the subinterval is $O_t(h^{-t-3})$, we have
\[
 s_1s_2\ge 2^{h+t},\qquad \sum_{i=1}^k s_i^{-1}<1,\qquad
 (s_k/s_1)^{t+1}<1+\binom{k}{t+1}^{-1}
\]
for all sufficiently large $h$.

If $2\le j<h$, then $L(J)\ge j(2x/3-1)\ge2^j$ for every $j$-set
$J$. To see the second inequality, it suffices to take $j=h-1$, since
$j/2^j$ decreases for $j\ge2$. In contrast, every $h$-set $J$
satisfies $L(J)\le 3hx/4<2^h$. Once an omitted $A$ with
$1-|A|\ne0$ has been found inside every $h$-set, each larger $J$ also
contains such an $A$, whether or not $L(J)<2^{|J|}$. Thus after the
compulsory singleton omissions, the remaining requirement is precisely
to meet condition~3 on all $h$-sets.

Let $u=h$ in the first case and $u=h-1$ in the second. In either case
$u$ is the largest integer at most $h$ such that $1-u\ne0$ in the
field. An additional omission in a minimizing family must occur in
some $h$-set and have nonzero $1-|A|$, for otherwise restoring the
order lowers the cost and preserves feasibility. Thus $|A|\le u$.
If $|A|<u$, replace $A$ by all $u$-sets $T$ containing it. Every
$h$-set previously met by $A$ contains at least one such $T$. Moreover,
\[
 \sum_{\substack{T\supset A\\|T|=u}}\frac{w(T)}{w(A)}
 =\sum_{\substack{I\subseteq[k]\setminus A\\|I|=u-|A|}}
          \prod_{i\in I}s_i^{-1}
 \le\left(\sum_{i\notin A}s_i^{-1}\right)^{u-|A|}<1.
\]
The family would have strictly smaller cost, even if no replacement
were already present. It follows that all additional omissions have
$|A|=u$.

If $u=h$, every $h$-set must itself be chosen. If $u=h-1$, record an
omission by its complement $B=[k]\setminus A$. The omitted order is
$n/p_A=p_B$ and its cost is $w(A)=c(B)$. For an $h$-set $J$ we have
$A\subseteq J$ exactly when $[k]\setminus J\subseteq B$.
Consequently the choices in the second case are precisely the coverings
described in the statement, with the indicated total costs. This also
explains the $t$-subsets in the first case by taking complements there.

It remains to recover the number of blocks from the dimension.
Set $b=t+1$, $N=\binom{k}{b}$, and $C=C(k,b,t)$. Any covering costs
at least $C s_1^b$, and one with $C$ blocks costs at most
$Cs_k^b<(C+1)s_1^b$ by our choice of the primes and $C\le N$.
A covering with more than $C$ blocks already costs at least
$(C+1)s_1^b$. Each covering of minimum total cost therefore has $C$
blocks, and dividing its cost by $s_1^b$ gives~\eqref{design:recover}.
\end{proof}

\begin{corollary}\label{gap:prescribe}
Given a nonempty finite set $\mathcal P$ of primes, there are infinitely
many squarefree lengths $n$ coprime to all its members for which
\[
 U(n)-M_q(n)=
 \begin{cases}
 p_1-1,&\lambda\in\mathcal P,\\
 1,&\lambda\notin\mathcal P,\quad \lambda\nmid n .
 \end{cases}
\]
Here $M_0(n)=U(n)-1$. One may take $k\to\infty$ with
$p_1\asymp 2^k/k$. In particular, the difference between the optimal
dimensions in any two prescribed distinct positive characteristics is
unbounded along a sequence of lengths admissible for both.
\end{corollary}

\begin{proof}
Apply Theorem~\ref{design:realization} with $t=0$ and
$h=k=1+(\prod_{\ell\in\mathcal P}\ell)^v$, letting $v$ tend to
infinity. In characteristic zero and in each admissible characteristic
outside $\mathcal P$, the additional omitted order is $p_\varnothing=1$.
For $\lambda\in\mathcal P$ we must cover the empty set by a singleton.
The least expensive choice is uniquely $\{1\}$, giving the omitted
order $p_1$. As $p_1\to\infty$, every member of $\mathcal P$ is
eventually admissible. For a pair of distinct characteristics, take
$\mathcal P$ to contain just the first. Both become admissible and the
difference in their optimal dimensions has absolute value $p_1-2$.
\end{proof}

As an example with $t=2$, let $k=2^a-1$, where $a$ is odd and tends to
infinity, and put $h=k-2$. Then $2\mid h-1$ and $3\nmid h-1$.
For the primes chosen in Theorem~\ref{design:realization}, all pair
orders $p_ip_j$ are additionally omitted in characteristic three.
In characteristic two the additional omissions are the triple orders
of a Steiner triple system of minimum total block cost. Indeed, after
identifying the $k$ points with the nonzero elements of $\mathbb F_2^a$,
the triples $\{x,y,x+y\}$ for distinct $x,y$ cover each pair exactly
once. Hence $C(k,3,2)=k(k-1)/6$ by counting pairs. Any covering of
this size must likewise contain each pair exactly once. The theorem
then selects precisely those Steiner triple systems for which the
sum of the corresponding costs is smallest.

\subsection{Squarefree lengths with short coordinate planes}

The plane condition in Theorem~\ref{design:covercriterion} can act
together with the characteristic condition. The following range gives
an explicit illustration. Assume $k\ge3$, write $L=L_k$, and set
$\alpha=\mathbf1_{\{p_1p_2<2^k\}}$.

\begin{corollary}\label{gap:region}
Suppose
\begin{equation}\label{gap:regionhyp}
 p_1p_2\ge 2^{k-1},\qquad L_h\ge2^h\quad(2\le h<k),\qquad
 L+\Delta\ge2^k .
\end{equation}
For $n\in S$ let $H_\lambda$ be the set of divisors $d$ of $n$ for
which $1-k+\omega(d)$ is nonzero in the coefficient field. Then
$d(D_S)=2^k$ if and only if
\begin{enumerate}
\item $n/p_i\notin S$ for every $i$,
\item $\Div(n/(p_ip_j))\not\subseteq S$ whenever $i<j$ and $p_ip_j<2^k$,
and
\item $H_\lambda\not\subseteq S$ whenever $L<2^k$.
\end{enumerate}
The maximizing spectrum is unique and
\[
 U(n)-M_q(n)=
 \begin{cases}
 \alpha,&L\ge2^k,\\
 1,&L<2^k,\quad 1-k\ne0,\\
 p_1-1+\alpha,&L<2^k,\quad 1-k=0,
 \end{cases}
\]
where $1-k$ is taken in the coefficient field.
\end{corollary}

\begin{proof}
Since $\Delta>0$, the hypotheses of Theorem~\ref{design:covercriterion}
hold. For a proper $J$ of cardinality $j\ge2$ we have
$L(J)\ge L_j\ge 2^j$. Hence only $J=[k]$ can occur in condition~3
of that theorem. In this case $1-|A|=1-k+\omega(n/p_A)$, and the
third condition takes the stated form.

We determine the omissions in addition to the $n/p_i$. If $L\ge2^k$,
only the planes with $p_ip_j<2^k$ need be excluded. There are such
planes precisely when $\alpha=1$, and omission of order one excludes
all of them. This is the unique omission of cost one.
If $L<2^k$ and $1-k\ne0$, then $1\in H_\lambda$, so the same omission
also meets the third condition and is the unique cheapest choice.

Suppose finally that $L<2^k$ and $1-k=0$. The order one is absent
from $H_\lambda$, whereas every prime order belongs to it. The unique
least expensive order in $H_\lambda$ is $p_1$, of cost $p_1-1$.
When $\alpha=1$, the plane with directions $1,2$ free has spectrum
$\Div(n/(p_1p_2))$. This spectrum contains neither $p_1$ nor any
compulsory omission. The plane thus requires a further omission, and
order one meets this and all remaining plane conditions. For uniqueness,
any order other than $1,p_1$ costs at least $p_2-1>p_1$, exceeding even
the combined cost of $1$ and $p_1$. This proves the formula in all cases.
\end{proof}

For example, take $m=7\cdot11\cdot13\cdot17\cdot19\cdot23\cdot29
=215\,656\,441$. Here $2^k=128$, $p_1p_2=77$, $L=112$ and
$\Delta=44$. The sums $L_2,\ldots,L_6$ are $16,28,44,62,84$.
For this squarefree $m$, the coordinate-line bound is
$U(m)=m-\sum_i\prod_{j\ne i}(p_j-1)=145\,501\,945$. Thus Corollary~\ref{gap:region} applies with
$\alpha=1$ and $L<128$. The short
planes require the omission of order one in every optimal spectrum.
In characteristics two and three, $1-k=-6$ vanishes and the order
seven must also be omitted. Its cost $\varphi(7)=6$ accounts for the
difference in the values
\[
 M_q(m)=
 \begin{cases}
 145\,501\,938,&\lambda\in\{2,3\},\\
 145\,501\,944,&\lambda\notin\{2,3\},\quad \lambda\nmid m.
 \end{cases}
\]

\section{The first three terms of the minimum redundancy}\label{sec:cubic}

The coordinate lines considered in Section~\ref{sec:saturation} each force the omission of at least one
order. We now consider faces with two or three free coordinates. Their contribution depends on how often
a word repeats in the remaining coordinates.

Write $n=\prod_{i=1}^kP_i$, where $P_i=p_i^{a_i}$, and identify an order $d_e=\prod_i p_i^{e_i}$ with its
exponent vector $e$. For $j\ge1$ define
\begin{equation}\label{cubic:scales}
 R_j(n)=\sum_{\substack{I\subseteq[k]\\ |I|=j}}\frac{n}{\prod_{i\in I}P_i},
\end{equation}
with an empty sum equal to zero. For each exponent vector put
\[
 z(e)=|\{i:e_i=0\}|,\qquad
 h(e)=\prod_{i:e_i>0}p_i^{a_i-e_i}.
\]
An index $e_i=0$ means that the corresponding word is constant in coordinate $i$. For $e_i>0$, the characters of order $p_i^{e_i}$ factor through the reduction modulo
$p_i^{e_i}$. Hence the local component is the $p_i^{a_i-e_i}$-fold periodic repetition of the top
component at length $p_i^{e_i}$. Its minimum distance is $2p_i^{a_i-e_i}$ by
Section~\ref{sec:prelim}. Define
\begin{equation}\label{cubic:terms}
 A_s(n)=\sum_{\substack{z(e)=s\\2^{s-1}\le h(e)<2^s}}\varphi(d_e),
 \qquad 1\le s\le3.
\end{equation}
In particular, $A_1(n)=\sum_i\varphi(n/P_i)$. To compare these quantities with the scales $R_s(n)$
of~\eqref{cubic:scales}, fix the set $I$ of zero indices of $e$. Each of the $2^{s-1}$ integers in the
indicated interval determines at most one possible $e$, by unique factorization of $h(e)$. Since
$\varphi(d_e)\le n/\prod_{i\in I}P_i$, we obtain $A_s(n)\le 2^{s-1}R_s(n)$.

\begin{theorem}\label{cubicmain}
Let $N_k=k+2\binom{k}{2}+4\binom{k}{3}$. If $\min_iP_i\ge14N_k$, then, over every admissible coefficient field,
\begin{equation}\label{cubic:mainbound}
 0\le n-M_q(n)-A_1(n)-A_2(n)-A_3(n)\le R_4(n).
\end{equation}
In particular, $|M_q(n)-M_{q'}(n)|\le R_4(n)$ for any two such fields. For fixed $k\ge3$ this difference
is $o(R_3(n))$ as $\min_iP_i\to\infty$.
\end{theorem}

\begin{proof}[Proof of the lower bound]
Let $I\subseteq[k]$ have size $s$, $1\le s\le3$. Consider a face of the exponent grid on which the
indices in $I$ are free and those outside $I$ are fixed at positive values $e_i$ with
\[
 2^{s-1}\le h=\prod_{i\notin I}p_i^{a_i-e_i}<2^s.
\]
If every order of this face is retained, its code contains the tensor product of a point indicator on
each free axis and a minimum word on each fixed axis. The resulting word has weight $2^{k-s}h<2^k$. Every
feasible spectrum must therefore omit an order on this face. Its least possible cost is $c=\prod_{i\notin
I}\varphi(p_i^{e_i})$, attained when all the free indices are zero. As the face is specified by $I$ and
$h$, there are at most $N_k$ such faces. Their least costs sum to $A_1+A_2+A_3$.

One omitted order may lie on several of these faces. We show that its cost accounts for the least costs
of all the faces containing it. Suppose two distinct faces through the same point have free sets $I$ and
$J$. If $I=J$, the fixed indices agree and so do the faces. If $I\subsetneq J$, the repetition factor of
the face with free set $J$ divides that of the face with free set $I$, whereas its defining interval
forces it to be strictly larger. Thus $I$ and $J$ are incomparable. On every face through a common point
one can consequently choose an axis $i$ which is free on that face and fixed on another, say at $u>0$.
Since the latter repetition factor is at most seven,
\[
 p_i^{a_i-u}\le7,\qquad \varphi(p_i^u)\ge P_i/14\ge N_k .
\]
If the point has cost $w$, the least cost of each incident face is at most $w/N_k$. The sum of these
costs is thus at most $w$. For a point on just one face the same conclusion follows directly from the
definition of least cost. Summing over omitted points counts every face at least once and proves the
lower bound in~\eqref{cubic:mainbound}.
\end{proof}

For the other inequality we shall construct a spectrum of the required dimension. We first give a lower
bound for the weight of a word in terms of its nonzero order components. In the following lemma $d(S)$
denotes the minimum distance of the code with order spectrum $S$.
\begin{lemma}\label{asy:margin}
For every nonempty spectrum $S$ on $j$ prime-power axes,
\begin{equation}\label{marginbound}
 d(S)\ge\min_{e\in S}2^{j-z(e)}h(e).
\end{equation}
Equality holds when every index of every $e\in S$ is positive.
\end{lemma}

\begin{proof}
First suppose that all indices are positive. On an axis of length $p^a$, let $v$ be the greatest index
occurring in a nonzero component of a given word $w$. Project the columns of $w$ onto level $v$ on this
axis, and choose a linear functional on that level which takes a nonzero value on one of the projected
columns. Applying the projection and then the functional to each column yields a nonzero word on the
other $j-1$ axes. Its orders belong to the section of $S$ at $v$. Each of its nonzero positions comes
from a different nonzero column of $w$. All these columns lie in the sum of levels $1,\ldots,v$, which is
the $p^{a-v}$-fold repetition of the length-$p^v$ zero-sum code. Hence each such column has weight at
least $2p^{a-v}$. Induction on $j$ now gives~\eqref{marginbound}, starting with a nonzero scalar on the
empty product. A tensor of minimum words in any component attaining the minimum has the stated weight.

For general $S$, take a nonzero word and choose an inclusion-maximal set $I$ of zero indices among its
nonzero order components. Sum the word over the position coordinates in $I$. The constant local component
survives summation, and every positive local component sums to zero. By maximality, the components which
survive have zero indices exactly in $I$. They still have distinct orders on the complementary axes, and
the scalar $\prod_{i\in I}P_i$ introduced by summation is invertible. The sum is thus nonzero, has only
positive indices on the remaining axes, and has weight no greater than the original word. The first part
gives the required lower bound.
\end{proof}

\begin{proposition}\label{cubic:construction}
Suppose $a_i\ge2$ whenever $p_i\in\{2,3\}$. Let $F$ consist of the orders $d_e$ for which
\begin{equation}\label{cubic:mask}
 z(e)\le3,\qquad h(e)\notin[2^{z(e)-1},2^{z(e)}).
\end{equation}
Then $d(F)=2^k$ over every admissible coefficient field.
\end{proposition}

We describe the maps used to reduce a prime-power axis. This description will also be used in the proof
of Proposition~\ref{gap:concrete}. For $1\le t\le a$, put $m=p^{a-t}$ and write the positions modulo
$p^a$ as $b+my$, where $0\le b<m$ and $y\in\mathbb Z/p^t\mathbb Z$. A word $w$ on this axis consists of
$m$ blocks $w_b(y)=w(b+my)$. Let $\sigma$ be a cyclic shift of the $p^t$ entries of a block. For
$\boldsymbol\tau=(\tau_b)_{0\le b<m}$, let
\[
 F_{\boldsymbol\tau}w=\sum_{b=0}^{m-1}\sigma^{\tau_b}w_b .
\]
We call $F_{\boldsymbol\tau}$ a folding map. Its value has weight at most $\wt(w)$, since each block is
permuted before addition. The joint kernel of these maps is
\begin{equation}\label{eq:foldkernel}
 \bigcap_{\boldsymbol\tau}\ker F_{\boldsymbol\tau}
 =\bigoplus_{u=1}^{a-t} V_{p^u}^{(p^a)} .
\end{equation}
Indeed, writing $\boldsymbol\varepsilon_b$ for the shift by one step in block $b$ alone gives
$(F_{\boldsymbol\varepsilon_b}-F_{\boldsymbol0})w=(\sigma-1)w_b$.
Thus a word in the joint kernel is constant, say $c_b$, in each block, and $F_{\boldsymbol0}w=0$ gives
$\sum_b c_b=0$. Conversely, these conditions imply $F_{\boldsymbol\tau}w=0$ for all $\boldsymbol\tau$.
Such words are the periodic lifts of the length-$m$ zero-sum code, which proves~\eqref{eq:foldkernel},
including the case $m=1$.

A constant word maps to a constant word. A character of order $p^{a-t+v}$, $1\le v\le t$, restricts on
block $b$ to a scalar multiple of a character of order $p^v$ in the variable $y$. Hence the component of
index $a-t+v$ maps into that of index $v$, also over the ground field. The joint map
$w\mapsto(F_{\boldsymbol\tau}w)_{\boldsymbol\tau}$ is injective on the sum of the constant and these $t$
components. In several coordinates the tensor product of the joint maps is injective on the tensor
product of these sums. Each other order component is killed in at least one direction. Distinct surviving
orders have distinct reduced exponent vectors. In particular, if a word has a nonzero component on the
retained orders, some choice of one folding map on each axis gives a nonzero word of no greater weight.

\begin{proof}[Proof of Proposition~\ref{cubic:construction}]
Take $t=2$ on axes with base $2$ or $3$ and $t=1$ on every other axis. On each retained order, folding
preserves $z(e)$ and $h(e)$, since $a-(a-t+v)=t-v$. Thus any nonzero image of the preceding maps
satisfies~\eqref{cubic:mask} on the reduced grid. If a nonzero word has no retained component, each of
its nonzero orders has a positive index at least $t$ below the top index on some axis. The corresponding
repetition factor is at least $p^t\ge4$. For $z\le2$ this gives $h\ge2^z$ directly, and for $z=3$
condition~\eqref{cubic:mask} excludes $4\le h<8$. In both cases Lemma~\ref{asy:margin} gives weight at
least $2^k$.

It remains to establish the bound on the reduced grid. To keep track of a section in which one or more
axes have been fixed at zero, let $\mathcal P$ be the set of integer pairs $(z,h)$
satisfying~\eqref{cubic:mask}. On any subproduct of the reduced axes put
\begin{equation}\label{cubic:shift}
 F_x=\{e:(z(e)+b,h(e))\in\mathcal P\text{ for all }0\le b\le x\},
 \qquad x\ge0 .
\end{equation}
Here and below $x,b$ are integers. In particular $F_x$ is empty for $x>3$. On a product of $j$ axes we
shall prove $d(F_x)\ge H=2^{j+x}$ for $0\le x\le3$. For subproducts of the axes of lengths $4,9,5,7$,
this is Lemma~\ref{cubic:core} of the appendix.

Adjoin a prime axis of length $p\ge11$ and argue by induction on the number of such axes. On the
preceding $j-1$ axes the section at exponent one is $F_x$. Its intersection with the section at exponent
zero is $F_{x+1}$, by~\eqref{cubic:shift}. Induction gives distance at least $H/2$ for the positive section and at least $H$ for the intersection
(with $F_4=\varnothing$ if $x=3$).
Write $s=3-x$. Slices along the new axis are words on the other $j-1$ axes. A difference of two slices
kills the constant component on the new axis, so belongs to the code of the section at exponent one.
Summing all slices kills the positive component, so the sum lies in the code of the section at zero.

If a slice is zero, every nonzero slice has weight at least $H/2$. Two such slices suffice. If exactly
one is nonzero, it is also the sum of the slices and thus belongs to the intersection of the two section
codes. Its weight is at least $H$. Now suppose all slices are nonzero. If two differ, the sum of their
weights is at least the weight of their difference, hence at least $H/2$. The spectrum of each slice is contained in the union of the two sections.
Every order in the positive section has at most $s$ zero indices on the remaining axes, and one in
the zero section has at most $s-1$. Lemma~\ref{asy:margin} therefore gives weight at least
$2^{j-1-s}$ for every other slice. The total is at least $H/2+(p-2)2^{j-1-s}\ge H$, since $p-2\ge2^{s+x}=8$. If all
slices coincide, their common value lies in the section at exponent zero because $p$ is invertible. This
section is empty for $s=0$. Otherwise its orders have at most $s-1$ zero indices on the remaining axes,
and the total weight is at least $p2^{j-s}\ge H$. This finishes the induction. Taking $x=0$ gives the
desired lower bound. The primitive component belongs to $F$ and contains a word of weight $2^k$, which
gives equality.
\end{proof}

\begin{proof}[Completion of the proof of Theorem~\ref{cubicmain}]
The assumption on $\min_iP_i$ ensures that bases $2,3$, when present, have exponent at least two. In the
construction just proved, the omitted orders with at most three zero indices have total cost
$A_1+A_2+A_3$. For a prescribed set $I$ of four zero indices, the sum of the dimensions of all orders
having zeros there is $n/\prod_{i\in I}P_i$. Summing over all four-subsets counts each omitted order
having at least four zeros at least once. Their total cost is therefore at most $R_4(n)$, which proves
the upper bound in~\eqref{cubic:mainbound}. The quantities $A_1,A_2,A_3$ do not depend on the coefficient
field, so two optimal dimensions differ by at most $R_4(n)$. Finally, $4R_4(n)\le(k-3)R_3(n)/\min_i P_i$
for $k\ge3$, by counting each four-subset via its four three-subsets. This gives the last assertion.
\end{proof}

\subsection{The two-term estimate at arbitrary lengths}

For $s=2$ the range in~\eqref{cubic:terms} consists of $h=2$ and $h=3$. One positive index must then lie
one below the top on the corresponding prime axis; the remaining positive indices are maximal. Thus
\[
 A_2(n)=
 \sum_{i<j}\ \sum_{\substack{\ell\notin\{i,j\}\\
              p_\ell\in\{2,3\},\ a_\ell\ge2}}
 \varphi\!\left(\frac{n}{P_iP_jp_\ell}\right).
\]
This expression permits a bound without the size assumption of Theorem~\ref{cubicmain}.
\begin{corollary}\label{asy:expansion}
Let $k\ge2$, $C_k=k+(k-1)(k-2)$ and $E_q(n)=n-M_q(n)-A_1(n)-A_2(n)$. Then
\[
 -30C_kR_3(n)\le E_q(n)\le R_3(n).
\]
If $\min_iP_i\ge5C_k$, or if both $4\nmid n$ and $9\nmid n$, the lower bound improves to $E_q(n)\ge0$.
When $k=1$ the minimum redundancy is one.
\end{corollary}

\begin{proof}
For the upper bound, keep the orders in~\eqref{cubic:mask} with $z(e)\le2$. If every base $2$ or $3$
present has exponent at least two, Proposition~\ref{cubic:construction} gives distance at least $2^k$.
Otherwise embed each exceptional axis of length $p\in\{2,3\}$ into the positions $0,p,\ldots,(p-1)p$ at
length $p^2$, extending words by zeros elsewhere. A zero-sum word on the embedded axis has zero sum in
every residue class modulo $p$, hence lies in the top component at length $p^2$. The image of a constant
word is constant on each residue class modulo $p$ and therefore lies in levels zero and one. In any
resulting order for which a zero index becomes positive, the number of zero indices falls to at most one
and $h$ acquires a factor of at least two. Such an order satisfies~\eqref{cubic:mask}. If every original
zero stays zero, both $z$ and $h$ are unchanged. Thus the simultaneous embeddings preserve weight and
send our code into the code of Proposition~\ref{cubic:construction}. The primitive component again gives
distance exactly $2^k$. The omitted orders with one or two zero indices cost $A_1+A_2$; those with at
least three cost at most $R_3(n)$.

For the lower bound we use the lines and planes in the proof of Theorem~\ref{cubicmain}. Denote by $F_i$
the line free in direction $i$ and maximal elsewhere, and by $F_{ij\ell}$ the plane free in $i,j$ with
index $a_\ell-1$ in direction $\ell$ and maximal elsewhere. The planes occur for $p_\ell\in\{2,3\}$ and
$a_\ell\ge2$. Each line and each plane must contain an omitted order. Their minimum costs sum to $A_1$
and $A_2$, respectively. If some $P_i$ is small, intersections can prevent us from adding all these
costs. We first choose faces for which the costs can still be added.

Put $W_i=\varphi(P_i)$. For the axes on which a plane can have submaximal index, put $m_i=W_i/p_i$; on
the other axes put $m_i=W_i$. Each fixed level on any of our faces has totient at least $m_i\ge P_i/5$.
Let $J=\{i:m_i<C_k\}$. Keep every line and all planes whose free pair contains $J$. There are at most
$k+2\binom{k-1}{2}=C_k$ retained faces. For each omitted point of cost $w$, sum the minimum costs of the
retained faces passing through it. We check that the sum is at most $w$.

Two distinct lines meet only at the retained top order. If two distinct planes meet, their free pairs
must differ: planes with the same free pair and different submaximal directions are disjoint. Each plane
through the point has a free index outside $J$ fixed by another plane, and hence minimum cost at most
$w/C_k$. The only line that can meet $F_{ij\ell}$ is $F_\ell$. Since $\ell\notin J$, such a line also
contributes at most $w/C_k$. Thus the bound holds at a point common to at least two planes. At a point
common to just one plane and one line, the two free indices of the plane are fixed on the line at their
top levels. The plane contributes $w/(W_iW_j)\le w/2$, because at most one $P_i$ has totient one. The
line contributes at most $w/C_k\le w/2$. A point lying on at most one retained face satisfies the bound
directly. Summing over the omitted points now bounds the redundancy below by the sum of all retained face
minima.

If $\min_i P_i\ge5C_k$, then $J$ is empty and every plane is retained. If $4\nmid n$ and $9\nmid n$,
there are no planes. In general, for each discarded plane choose $v\in J\setminus\{i,j\}$; then
$P_v<5C_k$. The minimum cost of this plane is at most
\[
 \frac{n}{P_iP_j}\le5C_k\frac{n}{P_iP_jP_v}.
\]
For each triple $\{i,j,v\}$ there are only three choices of free pair and at most two choices of $\ell$.
The discarded minima therefore sum to at most $30C_kR_3(n)$. This proves the claimed lower bounds. The
assertion for $k=1$ follows from Theorem~\ref{sat:all}.
\end{proof}

The scale $R_3(n)$ of the error in the two-term formula can be seen even when the number of axes is four.
Let $n=2^a11^b13^c17^d$ and take each odd prime power to be the largest power of its base not exceeding
$2^a$. For large $a$ the only order contributing to $A_3$ has index $a-2$ on the binary axis and zero
indices elsewhere: among powers of the four primes, only $4$ lies in $[4,8)$. Hence $A_3=2^{a-3}$. The
four prime powers tend to infinity, and $R_4(n)=1$, so Theorem~\ref{cubicmain} yields
\[
  2^{a-3}\le E_q(n)\le 2^{a-3}+1
\]
over every admissible field. Since $2^a\le R_3(n)\le4\cdot2^a$, the two-term remainder $E_q(n)$ has order $R_3(n)$ in
every admissible characteristic. Its leading contribution $A_3(n)$ is common to all these fields.
On these same lengths Theorem~\ref{cubicmain} bounds the difference between any two optimal dimensions
by $R_4(n)=1$.

\section{The dependence of the maximum dimension on the characteristic}
\label{sec:persistent}

Theorem~\ref{cubicmain} shows that, for a fixed number of prime
divisors, the first three terms of the minimum redundancy are common to
all admissible characteristics. We first give a family in which the
remaining term still distinguishes two characteristics, however large
the prime powers become. We then examine the effect of adjoining new
prime divisors to the length.

\subsection{Lengths with ten fixed prime divisors}

Take the ten primes
\[
 (p_1,\ldots,p_{10})=(67,71,73,79,83,89,97,101,103,107)
\]
and let $n=\prod_{i=1}^{10}P_i$, where $P_i=p_i^{a_i}$ and $a_i\ge3$.
For this choice, $2^6<p_i<2^7$ for each $i$ and $\sum_i p_i=870$.
Denote by $S_0$ the set of divisors of $n$ remaining after the following
orders have been omitted:
\begin{equation}\label{gap:faces}
 \frac{n}{P_i}\quad(i\in[10]),\qquad
 \frac{n}{p_j\prod_{i\in I}P_i}
 \quad (I\subseteq[10],\ |I|=7,\ j\notin I).
\end{equation}
Let $C=\sum_{d\in\Div(n)\setminus S_0}\varphi(d)$ be the dimension lost
by omitting these components.

\begin{proposition}\label{gap:concrete}
Let $\lambda$ denote either zero or a prime not dividing $n$. Then
\[
 M_\lambda(n)=n-C-\mathbf1_{\{\lambda=7\}}.
\]
The maximizing spectrum is unique. It equals $S_0$ when $\lambda\ne7$
and $S_0\setminus\{1\}$ when $\lambda=7$.
\end{proposition}

\begin{proof}
\emph{The possible spectra.}
Write $D=2^{10}$. Each of the ten coordinate lines through the top of
the exponent grid requires an omission. Given a pair $(I,j)$ occurring
in~\eqref{gap:faces}, consider also the face in which the seven indices
in $I$ are free, the index at $j$ is $a_j-1$, and the other two indices
are maximal. If all orders on this face were retained, we could take a
point indicator in each free coordinate and a minimum-weight word in
each of the three fixed components. Their tensor product would have
weight $8p_j<D$. Thus each of these faces also requires an omission.
There are $10+10\binom{9}{7}=370$ lines and faces in total. In each
one the unique point of least cost is obtained by setting every free
index equal to zero. Its order is the corresponding divisor
in~\eqref{gap:faces}.

We claim that a feasible spectrum of redundancy at most $C+1$ omits
all these least-cost points. Indeed, distinct intersecting faces have
incomparable free index sets, counting a line here as a face. Faces
of dimension seven with the same free set and different $j$ are
disjoint. A line whose free index belongs to $I$ cannot meet the face
$(I,j)$, since the two prescribed indices at $j$ differ. Hence at a
point shared by two or more faces, each incident face has a free
index which another face fixes at $a_i$ or $a_i-1$. The corresponding
totient factor is at least
$67\cdot66=4422$, since $a_i\ge3$. If the shared point has cost $b$,
the sum of the least costs of its incident faces is therefore at most
$370b/4422<b-1$. On a single face, a point other than its minimum
costs at least $66$ times the minimum. In particular, all the minima
are distinct. Assign to each omitted point the least cost of every
face containing it. Every face is counted at least once, so the
assigned costs sum to at least $C$. If a face fails to omit its
minimum, one of its omitted points costs more than its total
assignment plus one. The redundancy would exceed $C+1$, proving the
claim. The order $1$ alone costs one, as all the primes are odd.
Consequently the only candidates at these two redundancies are $S_0$
and $S_0\setminus\{1\}$.

\emph{The form of a word of weight below $D$.}
We shall need the same reduction for $S_0$ and for each of its
sections obtained by fixing one index at its maximum. Let $w$ be a
nonzero word in the corresponding code and let $t$ be the number of
its coordinates in the product grid, so that $t=10$ or $9$.
In each of these $t$ directions in turn, apply the block-sum maps
of~\eqref{eq:foldkernel} with target length $p_i$. If some map has a
nonzero image, choose such a map and pass to its image. Otherwise
the joint kernel consists of the intermediate levels $1,\ldots,a_i-1$.
Take the greatest occupied level $v_i$, choose a linear functional on
it which is nonzero on the corresponding component, and extend the
functional by zero on the other levels. Contraction on this axis
gives a nonzero array on the remaining axes. Each nonzero entry
comes from a distinct nonzero column in levels $1,\ldots,v_i$.
Such a column is a $p_i^{a_i-v_i}$-fold repetition of a zero-sum
word at length $p_i^{v_i}$ and thus weighs at least $2p_i^{a_i-v_i}$.

Let $u$ be the word obtained after all axes have been processed, and
let $j$ be the number of axes on which a block-sum map was chosen.
These $j$ axes now have prime lengths. Since taking a block sum can
only decrease weight, we have
\[
 \wt(w)\ \ge\ 2^{t-j}H\wt(u),\qquad
 H=\prod_{i\text{ contracted}}p_i^{a_i-v_i}.
\]
Suppose that $\wt(w)<2^t$. If a contraction occurs, then
$67^2>2^{10}$ forces exactly one contraction and $v_i=a_i-1$ there.
Thus $j=t-1$ and $\wt(u)<2^{t-1}/p_i<67$.

For any set $I$ of seven of these $j$ prime axes, the component of
$u$ with zero indices exactly in $I$ vanishes. Indeed, its only
possible source has original indices zero on $I$, index $a_i-1$ at
the contracted axis, and top indices at every other axis (including
the fixed axis when $t=9$). This is the second omission
in~\eqref{gap:faces}. Now sum $u$ over the seven axes in $I$. Its
projection to the top component of the remaining axes is zero by
the preceding observation. In each remaining direction $\ell$, let
$E'_\ell$ be the positions used by $u$. Since $|E'_\ell|\le\wt(u)<67$,
this is a proper subset of the $p_\ell$ positions. The projection
onto the zero-sum space has the constants as its kernel. It is
therefore injective on the span of the unit vectors indexed by
$E'_\ell$. The seven-axis sum belongs to the tensor product of these
spans, on which the product projection is still injective. Hence the
sum itself is zero. Summation over $I$ kills each component
nonconstant on one of its axes and multiplies the other components
by $\prod_{i'\in I}p_{i'}$, a unit. Applying this to every $I$ shows
that all components of $u$ have at most six zero indices.
By~\eqref{marginbound}, $\wt(u)\ge2^{j-6}$, which gives
$\wt(w)\ge2p_i2^{t-7}>2^t$. This excludes the contraction.

It follows that every word of weight below $2^t$ as above has a
nonzero image on the squarefree grid under block-sum maps alone.
This image has no component with exactly one zero index, by the
first omissions in~\eqref{gap:faces}. On any nine of the axes the
two smallest side lengths have product greater than $512$, and
$66h\ge2^h$ for $2\le h\le9$. Theorem~\ref{sat:all} gives distance
$512$ for the squarefree code omitting precisely the nine indicated
orders. Hence each of the original nine-axis top sections has distance
at least $512$. For $t=10$, if the image also had zero top component,
it would be a sum of functions of at most eight coordinates.
Lemma~\ref{gap:flat} would then give weight at least $p_1p_2>D$.
Only the original top component can contribute to the top component
of the squarefree image. We conclude that a nonzero $w\in D_{S_0}$
of weight below $D$ has nonzero top component.

Consider such a $w$. For each $i$, partition the positions of the
$i$th axis into residue classes modulo $p_i^{a_i-1}$, each containing
$p_i$ positions. Within one class, all levels below $a_i$ are
constant. A difference of two slices in that class therefore lies
in the nine-axis section at $a_i$. If the slices differ, their
weights sum to at least $512$. The nonzero top component requires a
class with differing slices on each axis, and $\wt(w)<1024$ permits
only one such class on that axis. Choose in it a slice of least
weight $b$. If two slices differed from the chosen slice, the class
would have weight at least $2(512-b)+(p_i-2)b\ge1024$.
Consequently the class has a unique exceptional position, which
we denote by $x_i$.

Put $x=(x_1,\ldots,x_{10})$. On each axis separate $x_i$ from the
other $p_i-1$ members of its class and leave all other classes intact.
The word is constant on the cells of the resulting product
partition. A cell with two nonsingleton factors already has more
than $D$ positions, so its value is zero. We can therefore write
\[
 w(y)=c\,\mathbf1_{\{x\}}(y)+
 \sum_{i=1}^{10} f_i(y_i)\prod_{s\ne i}\mathbf1_{\{x_s\}}(y_s),
\]
where $f_i$ is constant on every residue class, taking on the class
of $x_i$ its value at a member other than $x_i$. The terms containing
$f_i$ have zero top projection on axis $i$. Thus $c\ne0$.

\emph{The exceptional characteristic.}
Evaluate this expression at a character of order $n/P_i$ and divide
by its value at $x$. The character is trivial on axis $i$ and has top
order on all other axes. This gives $\sum_y f_i(y)=-c$ for every $i$.
Next take an order of the second kind in~\eqref{gap:faces}, with zero
indices on $I$ and index $a_j-1$ at $j$. In the same normalization the point
term contributes $c$, and the seven sums from $I$ contribute $-7c$.
The two top-order characters annihilate the corresponding blockwise
constant functions. Thus for every character $\chi$ of exact order
$p_j^{a_j-1}$ on the $j$th axis,
\[
 \widehat f_j(\chi)=6c\,\chi(x_j),\qquad
 \widehat f_j(\chi)=\sum_y f_j(y)\chi(y).
\]
If $f_j$ is supported on just one class, say the class of $z$, then
$\chi$ is constant there and $\widehat f_j(\chi)=-c\chi(z)$.
We obtain $a=\chi(z-x_j)=-6$. The order of $\chi$ is odd, so the
same equation holds for $\chi^2$ and gives $a^2=-6=a$.
Since $a\ne0$, we must have $a=1$, or $7=0$ in the coefficient field.
In every other characteristic all ten functions $f_j$ consequently
occupy at least two classes. Off $x$, the corresponding coordinate
lines are disjoint, and $\wt(w)\ge\sum_j(2p_j-1)>D$. We have proved
that $S_0$ preserves the distance whenever $\lambda\ne7$.

In characteristic seven let $L_i$ be the subgroup of order $p_i$
on the $i$th coordinate axis through the origin. The nonzero word
\[
 \mathbf1_{\{0\}}-\sum_{i=1}^{10}p_i^{-1}\mathbf1_{L_i}
\]
has weight at most $1+\sum_i(p_i-1)=861<D$. At a character with index vector $e$, the sum of the
character over $L_i$ is $p_i$ if $e_i<a_i$ and zero otherwise. The Fourier coefficient of the displayed
word is therefore $1-\#\{i:e_i<a_i\}$. The cardinality here is one or eight at
the omitted orders~\eqref{gap:faces}, so the word lies in $D_{S_0}$.
On the other hand, every nonzero word of $D_{S_0}$ of weight below
$D$ has the form just proved. Its coordinate sum is
$c+\sum_i\sum_y f_i(y)=-9c\ne0$ in characteristic seven.
Omitting order one therefore excludes all such words. The primitive
component still supplies a word of weight $D$. Together with the
first part of the proof, this gives both the maximum dimension and
its unique spectrum.
\end{proof}

\subsection{Adjoining a new prime divisor to the length}

Let $m\ge1$ and let $r$ be a prime not dividing $m$. Over any field
of characteristic zero or prime characteristic coprime to $mr$, denote its characteristic by
$\lambda$. Then
\begin{equation}\label{eq:prime-extension}
 (r-1)M_\lambda(m)\le M_\lambda(mr)\le(r-1)M_\lambda(m)+m.
\end{equation}
For the lower bound, take the tensor product of a maximizing code
at length $m$ with the zero-sum code $V_r^{(r)}$ of length $r$.
It contains $V_{mr}^{(mr)}$ and has distance $2^{\omega(m)+1}$.
For the upper bound, let $S$ be a maximizing spectrum at length $mr$
and set $B=\{d\mid m:dr\in S\}$. The code $D_B\otimes V_r^{(r)}$
is a subcode of $D_S$, so $2d(D_B)\ge2^{\omega(mr)}$. Since $m\in B$,
we also have $d(D_B)\le d(V_m^{(m)})=2^{\omega(m)}$. Hence $D_B$
is feasible at length $m$ and has dimension at most $M_\lambda(m)$.
The orders divisible by $r$ contribute at most $(r-1)M_\lambda(m)$ to
the dimension of $D_S$ and the remaining orders at most $m$.

For example, let the ten exponents in Proposition~\ref{gap:concrete}
tend to infinity, and for each resulting length $n$ take a prime
$r>n^2$. The eleven prime-power factors of $nr$ all tend to infinity.
Applying~\eqref{eq:prime-extension} in characteristics two and seven
yields
\[
 r-1-n\ \le\ M_2(nr)-M_7(nr)\ \le\ r-1+n .
\]
The difference is asymptotic to $r$, and hence also to
$R_{10}(nr)=r+\sum_iP_i$. Here it is the difference between two characteristics that has order
$R_{10}(nr)$. On these lengths $R_4(nr)\ge r(\min_iP_i)^6$, by taking a term in
\eqref{cubic:scales} that retains $r$. Thus $R_{10}(nr)=o(R_4(nr))$. Whether a characteristic
difference can attain the $R_4$ scale in Theorem~\ref{cubicmain} remains undetermined by these examples.

In general, the difference $M_\lambda(mr)-(r-1)M_\lambda(m)$ lies in $[0,m]$ by
\eqref{eq:prime-extension}. Subtracting this expression for the two characteristics introduces an
error of absolute value at most $m$. Thus for fixed $m$ and distinct primes $\lambda,\lambda'$ not
dividing $m$, division by $mr$ gives
\begin{equation}\label{eq:gap-limit}
 \lim_{\substack{r\to\infty,\ r\ {\rm prime}\\r\nmid m\lambda\lambda'}}
 \frac{M_\lambda(mr)-M_{\lambda'}(mr)}{mr}
 =\frac{M_\lambda(m)-M_{\lambda'}(m)}{m}.
\end{equation}
Here the number of prime divisors stays fixed at $\omega(m)+1$.
Taking $m=215\,656\,441$ and substituting the two dimensions computed after
Corollary~\ref{gap:region} gives a limit of $6/m$ for $\lambda=5$ and $\lambda'=2$.

We next allow the number of prime divisors to increase. Write
$\rho_q(n)=M_q(n)/n$ for the maximum rate among the cyclic
supercodes in~\eqref{eq:objective}.

\begin{corollary}\label{gap:rates}
For any two distinct primes $\lambda,\lambda'$ there exists a
sequence of squarefree lengths $n_j$ coprime to $\lambda\lambda'$
such that
\[
 0<\lim_{j\to\infty}\rho_{\lambda'}(n_j)
 <\lim_{j\to\infty}\rho_\lambda(n_j)<1 .
\]
The common minimum distances $d_j=2^{\omega(n_j)}$ tend to infinity,
and the relative distances $d_j/n_j$ tend to zero.
\end{corollary}

\begin{proof}
Apply Corollary~\ref{gap:prescribe} to $\mathcal P=\{\lambda'\}$ and choose a
squarefree length $m$ whose least prime factor $p_1$ exceeds $2,\lambda,\lambda'$.
Then $M_\lambda(m)-M_{\lambda'}(m)=p_1-2$. The two rates at $m$
both belong to $(0,1)$, since the primitive component is nonzero and
its distance is greater than one. We will preserve their limiting
values by adjoining sufficiently large primes.

For each $j\ge2$ choose $j$ distinct primes $r_{j,1},\ldots,r_{j,j}$
greater than $j^2$ and prime to $m\lambda\lambda'$, and put
$n_j=m\prod_{i=1}^j r_{j,i}$. For either characteristic $\mu$, divide
\eqref{eq:prime-extension} at lengths $a,ar$ by $ar$. As
$0\le\rho_\mu(a)\le1$, a single step changes the rate by at most
$1/r$. It follows that
\[
 |\rho_\mu(n_j)-\rho_\mu(m)|
 \le\sum_{i=1}^j\frac{1}{r_{j,i}}<\frac1j .
\]
The two limiting rates are thus their respective values at $m$, with
difference $(p_1-2)/m$. Finally $\omega(n_j)=\omega(m)+j$, so
$d_j\to\infty$, while
\[
 \frac{d_j}{n_j}
 =\frac{2^{\omega(m)}}{m}\prod_{i=1}^{j}\frac{2}{r_{j,i}}
 \le\frac{2^{\omega(m)}}{m}\left(\frac{2}{j^2}\right)^j
 \longrightarrow 0 .
\]
\end{proof}

\appendices

\section{Distances on subproducts of the four-coordinate grid}\label{app:cubic}

The reduction in the proof of Proposition~\ref{cubic:construction} leaves
four possible side lengths below eleven. We determine the lower bounds
needed to start the induction on the number of prime coordinates.

\begin{lemma}\label{cubic:core}
Let $G$ be the product of any $j$ of the four coordinate sets of sizes
$4,9,5,7$. Let the coefficient field have characteristic zero or
characteristic prime to the product of the chosen sizes. For the spectra
$F_x$ in~\eqref{cubic:shift},
\[
                    d(F_x)\ge 2^{j+x}\qquad(0\le x\le3).
\]
\end{lemma}

We give the proof in three steps. Some elementary observations will be used
in each of them. For an exponent vector on these four axes, the possible
values of the repetition factor $h$ are $1,2,3,6$. Substitution
in~\eqref{cubic:mask} gives
\[
\begin{array}{c|cccc}
 h&1&2&3&6\\\hline
 \text{permitted values of }z\text{ in }F_0&
 \{0,2,3\}&\{0,1,3\}&\{0,1,3\}&\{0,1,2\}.
\end{array}
\]
For membership in $F_x$, each of $z,z+1,\ldots,z+x$ must occur in the
indicated entry.

Suppose that the coordinates are divided into two groups. Index the
nonempty sections in the second group by the corresponding levels
$L_1,\ldots,L_s$ in the first group. We order the levels so that the
available section-distance bounds $b_1,\ldots,b_s$ are nondecreasing.
Let $a_i$ be a lower bound on the distance of the code on the first
group with levels $L_1,\ldots,L_i$. Then
\begin{equation}\label{app:sectionbound}
                         d\ge\min_{1\le i\le s}a_i b_i .
\end{equation}
Indeed, let $L_i$ be the last level occurring in a nonzero word. Choose a
linear functional on the first group which vanishes on the earlier
levels and gives a nonzero contraction on the $L_i$-component. The
contracted word belongs to the section at $L_i$. Each of its at least
$b_i$ nonzero positions comes from a different column of weight at
least $a_i$.

For a single axis $p^a$, consider a nonempty set of levels with largest
index $v$. If it contains every index from $0$ to $v$, its distance is
$p^{a-v}$. If some index is missing, then $v\ge1$ and the distance is $2p^{a-v}$. To see this, remove
the $p^{a-v}$ repetitions. At length $p^v$ the first code is the full
space. The second is a proper cyclic code, so contains no unit vector.
Its level $v$ contains a difference of two unit vectors.

\subsection{Two prime-square coordinates}

On the $4\times9$ grid we abbreviate the exponent pair $(i,j)$ to $ij$.
Put
\[
\begin{aligned}
 A&=\{22,12,21,11,01,10,00\},\\
 B&=\{12,21,11,02,20,00\},\\
 C&=\{22,11,01,02,10,20\}.
\end{aligned}
\]
The four row indices form two pairs according to their residue modulo two.
The nine column indices form three triples according to their residue
modulo three. A pair and a triple determine a $2\times3$ block.

We first prove $d(A)\ge4$. Since $20$ and $02$ are missing from $A$,
the two row sums in each pair are equal and the three column sums in
each triple are equal. Suppose that a nonzero word has weight at most
three. If all column sums vanish, every occupied column has at least
two entries, so only one column is occupied. The row-sum condition then
pairs its nonzero entries into equal pairs. Only one pair can occur,
and its sum is nonzero. If some column sum is nonzero, the three
columns in its triple each contain exactly one entry, with the same
value in all three. In each row pair the counts of these entries must
agree in the field. The counts lie between zero and three, and the
characteristic is zero or at least five. They therefore agree as
integers, which would make the total number of entries even.

To prove $d(B)\ge4$, use the omission of $22$. Within each block the
entries have the form $a_i+b_j$, since the component with zero row
and column sums is absent. Such a nonzero block has weight at least
two. A word of weight at most three thus occupies just one block.
Missing level $10$ makes the totals on the two row-parity classes equal.
One class is empty, so the sum over the occupied block is zero.
If its two rows agree, the weight bound permits only one nonzero
column, with two equal entries. If they differ, the row difference
is a fixed $c\ne0$, and each of the three columns must contain precisely
one nonzero entry equal to $c$ or $-c$. The block sum is then one of
$\pm c,\pm3c$. In both cases the sum is nonzero, a contradiction.

Let $T_i$ denote the first $i$ terms of the list
\[
                      11,\quad00,\quad12,\quad21,\quad02,\quad20,\quad22 .
\]
Use the row coordinate as the first group in~\eqref{app:sectionbound}. The second column below gives
the order in which the row levels are taken. For each prefix of this list, the single-axis formula
gives its distance $a_j$ at length four. At length nine the same formula gives the distance $b_j$ of
the section of $T_i$ at the $j$th listed row level. Substitution in~\eqref{app:sectionbound} gives the
last column. For $i=6$ we use the sharper bound already proved for $B$.
\[
\begin{array}{c|c|c|c|c}
 i & \text{row levels} & (a_j) & (b_j) & d(T_i)\text{ is at least}\\\hline
 1 & (1)     & (4)     & (6)     & 24\\
 2 & (1,0)   & (4,2)   & (6,9)   & 18\\
 3 & (1,0)   & (4,2)   & (2,9)   & 8\\
 4 & (1,2,0) & (4,2,1) & (2,6,9) & 8\\
 5 & (1,0,2) & (4,2,1) & (2,2,6) & 4\\
 6 & \multicolumn{3}{c|}{T_6=B,\ \text{by the preceding block argument}} & 4\\
 7 & (2,1,0) & (2,2,1) & (1,2,2) & 2
\end{array}
\]
For example, $T_4=A\cap B$. Its column sections at row levels $1,2,0$ are $\{1,2\}$, $\{1\}$ and
$\{0\}$, respectively. Hence $d(A\cap B)\ge\min\{4\cdot2,2\cdot6,1\cdot9\}=8$.
The layer with indices $12$ supplies a tensor of weight $4\cdot2$, so $d(A\cap B)=8$.

We shall also use four distances on the $5\times7$ grid. The code with
levels $01,10$ consists of the functions $f(i)+g(j)$ with both $f$ and
$g$ of total sum zero. If just one summand varies, its nonzero entries
are repeated on the other axis, giving weight at least ten. If both
vary, choose two rows with unequal $f$-values. Each column contributes
to at least one of these rows, and each of the remaining three rows
contains a nonzero entry. This again gives weight at least ten.
A difference in the length-seven direction attains ten. The spectra
$\{01,10,11\}$, $\{11\}$, $\{00\}$ have distances $2,4,35$,
respectively, by the zero-sum condition on the whole grid, the tensor
product formula, and repetition.

\subsection{Proper subproducts}

On the empty product only $F_0$ is nonempty and has distance one. On an
axis of size $p^2$ with $p=2,3$ the distances of $F_0,F_1$ are $2,2p$.
On a prime axis of size five or seven they are $2,\infty$.
In either case $F_x$ is empty for $x\ge2$.

On a pair of axes of sizes $4,9$ we have $F_0=A$. Every other pair
contains a prime axis of size at least five. A word of weight less
than four then has an empty slice in this direction. Differences of
slices belong to the positive-level section, so each nonempty slice
lies in the one-axis $F_0$. If there is only one such slice, it also
belongs to the level-zero section, since the sum of the slices lies
there. The intersection is the one-axis $F_1$. The single-axis
distances consequently give $d(F_0)\ge4$ for each pair. For larger
values of $x$ the bounds are as follows.
\[
\begin{array}{c|ccc}
 \text{side lengths}&d(F_1)\ge&d(F_2)\ge&d(F_3)\\\hline
 4,9&8&24&\infty\\
 p^2,r\quad(p=2,3,\ r=5,7)&4p&\infty&\infty\\
 5,7&35&\infty&\infty .
\end{array}
\]
In the first row $F_1=T_4$ and $F_2=\{11\}$. In the second row
$F_1=\{00,11\}$, and ordering the row levels as $1,0$ produces
the bounds $4p,pr$ in~\eqref{app:sectionbound}. Only the constant layer
occurs in $F_1$ in the last row.

We next treat three axes. On $p^2,5,7$, the sections of $F_1$ at the
levels $1,0,2$ of the first axis are $\{11\}$, $\{01,10\}$, $\{00\}$
on $5,7$. Their distances are $4,10,35$, and the prefix distances on
the first axis are $2p,p,1$. This gives $d(F_1)\ge16$. Both $F_2$ and
$F_3$ are empty. On $4,9,r$ with $r=5,7$, the first six terms defining
$T_i$ have section distances $1,2,2,2,r,r$ in $F_1$ on the final
axis. The table for $T_i$ gives $d(F_1)\ge16$. Here
$F_2=\{(1,1,1)\}$ has distance $48$, and $F_3$ is empty.

It remains to show $d(F_0)\ge8$ for these triples. If a side has size
seven, slice in this direction. An empty slice forces every other slice
into the two-axis $F_0$, and a single nonempty slice lies in $F_1$.
The pair bounds rule out a nonzero word of weight below eight in this
case. If all seven slices are nonzero, the only remaining possibility
is seven slices of weight one. The difference of two such slices has
weight at most two and lies in a positive-level section of distance
at least four. All seven therefore coincide. Their sum belongs to the
zero-level section, which omits the highest order on the other two axes.
A unit vector has a nonzero projection at this order, giving a
contradiction.

For the axes of sizes $4,9,5$, the positive-level and zero-level sections
in the last direction are $A$ and $B$. The preceding argument for an
empty slice applies with $d(A)\ge4$ and $d(A\cap B)=8$. Suppose then
that all five slices are nonzero and that their total weight is at most
seven. One is a word $E$ of weight one. Every different slice has
weight at least three, since its difference with $E$ lies in $D_A$.
At least four slices equal $E$, and all five cannot agree because the
zero-level section $B$ omits $22$. Translate and scale so that $E$ is
the unit vector at the origin. Write $v$ for the remaining slice. We have
\[
                  h=v-E\in D_A,\qquad g=v+4E\in D_B .
\]
Since $h\ne0$ and $d(A)\ge4$, the word $v$ has weight three and avoids
the origin. Hence $g(0,0)=4$ and $\wt(g)=4$.

Each occupied $2\times3$ block of $g$ is additive and has weight at
least two. If there are two occupied blocks, each has weight two and
therefore consists of one column of two equal nonzero entries.
Since $01$ is absent from $B$, the totals on the three column-residue
classes agree. At least one class is empty, so these totals vanish.
The two blocks must belong to the same column class since their
individual sums are nonzero. They then occupy different row-parity
classes. The omission of $10$ says that their totals agree, whereas
their sum is zero. As $2$ is invertible, both would have total zero.

Thus $g$ is supported in the block of the origin, and its total sum is
zero by the omission of $10$. The same block contains $h=g-5E$.
Since $20,02$ are omitted from $A$, the row sums of $h$ agree, as do
its column sums. Its total is $-5$, so the origin row and origin column
of $g$ have sums $5/2$ and $10/3$. In an additive $2\times3$ block
with total zero, an entry is the sum of its row and column means.
It follows that
\[
                   g(0,0)=\frac{5}{6}+\frac{5}{3}=\frac{5}{2}.
\]
This is different from $4$ in every characteristic admissible here.
This proves the assertion for three axes.

\subsection{The four-coordinate product}

For $F_1$ the seven $4,9$ indices in the list defining $T_i$ have sections
on $5,7$ given, in order, by $\{01,10,11\}$ once, $\{11\}$ three
times, $\{01,10\}$ twice, and $\{00\}$ once. Their distances are
$2,4,4,4,10,10,35$. Using the prefix bounds from the table
in~\eqref{app:sectionbound} gives $d(F_1)\ge32$. In $F_2$ only the
index $11$ occurs on $4,9$, paired with the top layer on $5,7$. Hence
$d(F_2)=24\cdot4=96$. Finally $F_3$ is empty.

We prove $d(F_0)\ge16$. In the two-coordinate section on $4,9$,
the spectrum is $A$, $B$, or $C$ according as zero, one, or two indices
of the $5,7$ coordinates vanish. In the length-seven direction the
positive section has distance at least eight, and its intersection with the
zero section has distance at least sixteen, by the result on $4,9,5$.
Consequently a nonzero word of weight at most fifteen cannot have an
empty slice in this direction. Neither can its seven slices all agree.
In that case their common value would lie in the zero section and have
weight at most two. It would have an empty length-five slice, forcing
each of its length-five slices into $D_B$, which has distance at least
four.

Take now a word of weight at most fifteen with unequal nonempty slices.
Any two unequal length-seven slices have combined weight at least eight.
If all slices had weight at least two, the five outside such a pair
would already add at least ten. Thus one slice has weight one; let $E$
denote it. Each slice different from $E$ has weight at least seven.
It follows that exactly six slices equal $E$ and the remaining slice $v$
weighs at most nine. Normalize $E$ to be the unit vector at the origin
of $4,9,5$, and write $E_0$ for the unit vector at the origin on $4,9$.
The slice difference $v-E$ belongs to the positive section, and the sum
$v+6E$ belongs to the zero section in the length-seven direction.

Write $V$ for the length-five slice of $v$ at zero. The other four
length-five slices of $v$ have all pairwise differences in $D_{A\cap B}$:
their differences can be taken in either $v-E$ or $v+6E$.
If they do not all agree, an unequal pair has combined weight at least
eight. As $\wt(v)\le9$, at least one of the four must then be empty.
Comparing with this empty slice, each nonempty one has weight at least
eight. There is just one, and $\wt(V)\le1$. Comparing the origin slice
with an empty slice, first in $v-E$ and then in $v+6E$, gives
\[
                    V-E_0\in D_A,\qquad V+6E_0\in D_B .
\]
Both words have weight at most two and must vanish. This would imply
$7E_0=0$.

It remains to consider the case where the four non-origin slices of $v$
all have the same value $W$. Take slice differences and sums in $v-E$
and $v+6E$. The resulting conditions are
\[
\begin{array}{ll}
 V-W-E_0\in D_A, & V+4W-E_0\in D_B,\\
 V-W+6E_0\in D_B, & V+4W+6E_0\in D_C .
\end{array}
\]
In particular $7E_0-5W\in D_B$. Since $4\wt(W)+\wt(V)\le9$,
we have $\wt(W)\le2$. The last word has weight at most three and
therefore vanishes. Thus $W=(7/5)E_0$ and $\wt(V)\le5$.

Let $\pi_{ij}$ denote projection onto the layer $ij$ on $4,9$.
Using $20\notin A$, $21\notin C$ and $22\notin B$ in the four
conditions gives
\[
 \pi_{20}V=\tfrac{12}{5}\pi_{20}E_0,\qquad
 \pi_{21}V=-\tfrac{58}{5}\pi_{21}E_0,\qquad
 \pi_{22}V=-\tfrac{23}{5}\pi_{22}E_0 .
\]
For a column $y\not\equiv0\pmod3$, the entries at $y$ of the length-nine
origin indicator projected onto levels $0,1,2$ are $1/9,-1/9,0$.
Indeed, the sum of the first two projections is constant on the
residue classes modulo three and vanishes off the class of zero.
In the length-four coordinate, taking the difference of the entries at
$0$ and $2$ kills levels $0,1$ and gives one on the top projection of
the origin indicator. We conclude that for all six such columns
\[
                 V(0,y)-V(2,y)=\frac{12}{45}+\frac{58}{45}
                                 =\frac{14}{9}\ne0 .
\]
Thus $V$ has weight at least six, contradicting the preceding bound.
This completes the proof of Lemma~\ref{cubic:core}.\hfill$\square$

\section*{Acknowledgment}
The authors used OpenAI's ChatGPT to draft and revise the abstract, Sections~\ref{sec:intro}--\ref{sec:persistent},
and the appendix. Its use included organizing the discussion of prior work, developing and checking
intermediate steps in the proofs, and checking parameter calculations and the distance estimates in
the appendix.

\section*{Data availability}
No external datasets were used in this work.

\section*{Declaration of competing interest}
The authors declare no competing interests.

\end{document}